\documentclass[preprint,aps,showpacs,floatfix]{revtex4}
\usepackage{graphicx}
\usepackage{epstopdf}

\usepackage[utf8x]{inputenc}
\usepackage{hyperref}
\usepackage{color}
\usepackage[dvipsnames]{xcolor}
\usepackage{footnote}
\usepackage{subcaption}
\usepackage{physics}

\usepackage{amssymb}

\usepackage{ulem}

\usepackage{dsfont}

\begin{document}
 \title{
 
 
 Measurement-based quantum Otto engine with a two-spin system coupled by anisotropic interaction: enhanced efficiency at finite times

 }
\author{Chayan Purkait}
 \email{2018phz0001@iitrpr.ac.in}
\author{Asoka Biswas}%
\affiliation{Department of Physics, Indian Institute of Technology Ropar, Rupnagar, Punjab 140001, India
}
\date{\today}%

\begin{abstract}
We have studied the performance of a measurement-based quantum Otto engine (QOE) in a working system of two spins coupled by Heisenberg anisotropic interaction. A non-selective quantum measurement fuels the engine. We have calculated thermodynamic quantities of the cycle in terms of the transition probabilities between the instantaneous energy eigenstates, and also between the instantaneous energy eigenstates and the basis states of the measurement, when the unitary stages of the cycle operate for a finite time $\tau$. 
The efficiency attains a large value in the limit of $\tau \rightarrow 0$ and then gradually reaches the adiabatic value in a long time limit $\tau \rightarrow \infty$. For finite values of $\tau$ and for anisotropic interaction, an oscillatory behaviour of the efficiency of the engine is observed. 
This oscillation can be interpreted in terms of interference between the relevant transition amplitudes 
in the unitary stages of the engine cycle. Therefore, for a suitable choice of timing of the unitary processes in the short time regime, the engine can have a higher work output and less heat absorption, such that it works more efficiently than a quasi-static engine. In the case of an always-on heat bath, in a very short time the bath has a negligible effect on its performance.

\end{abstract}

\pacs{}%
\maketitle


\section{Introduction}
The laws of classical thermodynamics are known to be applicable to the thermodynamic limit. It is quite interesting to study whether these laws are valid also in the quantum limit. 
In this regard, thermal machines (e.g. heat engines and refrigerators) can be considered as a suitable platform to explore this issue in quantum systems. Deviations from the classical limit of efficiency of these machines can be an important marker to understand the effect of quantum mechanical properties of the system.

In fact, it is rather crucial to explore whether it is possible to enhance the efficiency of thermal machines by harnessing quantum features such as coherence, many-body correlations, and non-thermal population distributions.
There have been several studies in different types of quantum heat engine (QHE) models to show that quantum coherence is indeed beneficial to achieve an enhanced performance of the QHEs\cite{latune2021EPJST,mitchison2015NJP,scully2003Science,shi2020JPA,latune2019SciRep,brandner2015NJP,uzdin2015PRX,uzdin2016PRAp,brandner2017PRL,scully2011PNAS,rahav2012PRA}. Roles of quantum correlation and entanglement  \cite{brunner2014PRE,Altintas2014PRE,barrios2017PRA,altintas2015PRA,hewgill2018PRA,zhang2007PRA}, the interaction within a coupled system  \cite{das2019Entropy,thomas2011PRE,ccakmak2016EPJP,ccakmak2017EPJP,altintas2015PRE,ivanchenko2015PRE,huang2020QIP,huang2014EPJP}, and the non-thermal heat baths \cite{de2020PRR,huang2012PRE,rossnagel2014PRL,alicki2015NJP,de2019PRL,scully2001PRL,scully2003Science} in the performance of quantum thermal machines have been investigated. It was shown that the efficiency of QHEs can be improved beyond the Carnot limit using squeezed thermal baths \cite{huang2012PRE,rossnagel2014PRL,klaers2017PRX}. But, their efficiency is bounded by a generalised efficiency limit for QHEs energised by non-thermal heat baths \cite{niedenzu2018NCom}.

From the time of Maxwell, it was known that work could be extracted from a single-temperature heat bath using information gained from measurements. This type of engine is known as Szilard's engine, in which results of selective measurement are used to provide feedback on engine operation \cite{maruyama2009RMP,liAP2012,jordanQSMF2019}. 
Recently, it was also shown that projective measurement of the ground state can be used to mimic the release of heat from a system to a cold bath during an isochoric process \cite{chand2017EPL,chand2017PRE,chand2021PRE} in an ion-based QHE.  
In later works, quantum measurement has been used to fuel the working system in a QHE, in which the isochoric heating stage in a standard quantum Otto engine (QOE) is replaced by a non-selective quantum measurement \cite{yiPRE2017,jordanQSMF2019,das2019Entropy,huangQIP2020}. Therefore, the engine works with a single heat bath as a heat shrink and non-selective quantum measurement as a heat source.

A finite-time analysis is also an important aspect of studying QHE, as for practical applications we need a finite amount of power. Moreover, a QHE in finite time may show true quantum nature in its performance which may not possible to observe in the quasistatic performance. Standard QHEs are operated by Hamiltonians who do not commute at different times  \cite{denzler2020PRR,dann2020Entropy,rezek2006NJP,lee2020PRE,camati2019PRA,ccakmak2017EPJD,turkpencce2019QIP,ccakmak2019PRE,plastina2014PRL,rezek2010Ent}. Consequently, quantum internal friction arises when a quantum system is driven unitarily  by an external control parameter in finite time. This induces nonadiabatic transitions between the instantaneous eigenstates of the Hamiltonian, and also generates coherence in the energy eigenbasis. As a result, a larger amount of entropy is produced and irreversibility is increased in engine operation, which degrades the performance of QHEs \cite{ccakmak2017EPJD,turkpencce2019QIP,ccakmak2019PRE,plastina2014PRL,rezek2010Ent}. On the other hand, in the presence of quantum coherence, QHEs can produce more power output than the classical ones  \cite{uzdin2015PRX,uzdin2016PRAp,brandner2017PRL,scully2011PNAS,rahav2012PRA,dodonov2018JPA}. Power output can be improved by not only the finite-time unitary stages, but also via the non-Markovian effects during finite-time bath interaction \cite{abiuso2019PRA}. In fact, in a very recent study on a finite-time QOE, it has been shown that coherence can act like a dynamical quantum lubricant \cite{camati2019PRA}. This can be interpreted as a dynamical interference effect, which takes place between the residual coherence after incomplete thermalization and the coherence generated in the subsequent finite-time unitary driving process. 

While the measurement processes and finite-time operation can individually have substantial effects on the performance of the QHEs, there have been very few studies when both protocols are used together. It is recently shown that it is possible to improve the performance of a single-qubit QHE by suitably choosing the measurement basis such that the degradation effect due to coherence production in a standard QOE can be overcome \cite{lin2021PRA}. In this work, we will investigate the finite-time performance of a two-spin QOE by using a non-selective quantum measurement to fuel the engine. 

In our model, two spins are coupled with each other by Heisenberg anisotropic XY interaction, in presence of external time-varying homogeneous magnetic fields. By changing the anisotropy parameter one can have the Heisenberg XX or Ising spin Hamiltonian as a limiting case. The free part and the driving part of the Hamiltonian do not commute, leading to the non-commuting nature of the Hamiltonian at two different times. Consequently, it initiates transitions between the instantaneous energy eigenstates, and introduces quantum features into the performance of a QHE through unitary driving processes in finite time. In this paper, we aim to investigate if a measurement-based engine operating in finite time performs better than when operated quasistatically and we show that it is indeed so. 
The corresponding enhancement of engine efficiency can be attributed to the anisotropy in the system and the use of the measurement protocol. 
Such an enhancement could not be achieved, if the engine was fueled by a heat bath, instead of a non-selective measurement.  We will also show that even if the spins remain coupled with a heat bath throughout the cycle (including the stages, when the magnetic fields are varied), it has a negligible effect on the engine's performance, for faster unitary stages. However, for a longer duration of these stages, coupling to the bath dominates and the performance of the engine degrades.

The paper is organized as follows. In Sec. \ref{QHE}, we introduce the two-spin model of the working system. We describe different stages of the quantum Otto cycle and the relevant thermodynamic quantities. Next in Sec. \ref{finite time operation}, we describe the finite time performance of the cycle. We provide a theoretical analysis of the thermodynamic quantities in terms of the transition probabilities. We also compare them with the quasistatic and the sudden limit of work and efficiency. In Sec. \ref{all time bath on}, we study the case when the thermal bath continuously interacts with the spins, even when the magnetic field is changed. We conclude the paper in Sec. \ref{concl}.


\section{Quantum heat engine model}\label{QHE}
In this section, we will briefly introduce our model of the QOE.

\subsection{System model}\label{model}

We consider a system of two spins coupled by Heisenberg anisotropic XY interaction in a transverse magnetic field [$B(t) \geq 0$]. The Hamiltonian is represented by \cite{ccakmak2019PRE, cherubim2022PRE, suzuki2012book}

\begin{equation}\label{Ham1}
    \hat{H(t)}=\hat{H_{0}}(t) + \hat{H_I},
\end{equation}
where,
\begin{eqnarray}\label{Ham2}
\hat{H_{0}}&=& B(t) (\hat{\sigma}_{1}^{z}+\hat{\sigma}_{2}^{z}) \\ \nonumber
\hat{H_{I}}&=&J[(1 + \gamma)~\hat{\sigma}_{1}^{x} \hat{\sigma}_{2}^{x} + (1 - \gamma)~\hat{\sigma}_{1}^{y}\hat{\sigma}_{2}^{y}].
\end{eqnarray}
Here $\hat{H}_{0}$ is the free part of the Hamiltonian, and $\hat{H}_{I}$ represents the interaction between two spins with $\gamma \in [-1,1]$ \cite{kamta2002PRL,yeo2005JPA} as the anisotropy parameter, and $J$ is the coupling constant between the spins. 
The operators 
$\hat{\sigma}_i^{x,y,z}$ are the standard Pauli matrices for the $i$th ($i\in 1,2$) spin. If $\gamma = 0$, the Hamiltonian becomes Heisenberg isotropic XX type, and for $\gamma = \pm1$, this becomes the Ising spin Hamiltonian. For $\gamma \neq 0$, $\hat{H}_{I}$ and $\hat{H}_{0}$ do not commute, which in turn gives rise to $[\hat{H}(t),\hat{H}(t')] \neq 0$. This further indicates that we may see quantum behaviour in finite-time engine operation \cite{rezek2010Entropy}.

The eigenvectors and the corresponding eigenvalues of the Hamiltonian $\hat{H}(t)$ are given by
\begin{equation}
  \begin{array}{l}
  \ket{\psi_{0}} = \frac{1}{\sqrt{2}}(\frac{B - K}{\sqrt{K^{2}  - BK}}\ket{11} + \frac{\gamma J}{\sqrt{K^{2}  - BK}} \ket{00}), ~E_{0} = -2K\\
  \ket{\psi_{1}} = \frac{1}{\sqrt{2}}(-\ket{10} + \ket{01}), ~~~~~~~~~~~~~~~~~~~~~E_{1} = -2J\\
  \ket{\psi_{2}} = \frac{1}{\sqrt{2}}(\ket{10} + \ket{01}), ~~~~~~~~~~~~~~~~~~~~~~~~E_{2} = 2J\\
  \ket{\psi_{3}} = \frac{1}{\sqrt{2}}(\frac{B + K}{\sqrt{K^{2}  + BK}}\ket{11} + \frac{\gamma J}{\sqrt{K^{2}  + BK}} \ket{00}),~ E_{3} = 2K
  \end{array}  
  \label{eigen}
\end{equation}
where $K=\sqrt{B^{2}+\gamma^{2}J^{2}}$. These energy eigenstates can be divided into two categories. The state that are dependent on the system parameters $B(t)$ and $J$, namely, $\left|\psi_{0}\right\rangle$ and  $\left|\psi_{3}\right\rangle$, evolve with time. The other ones which are independent of the system parameters, namely $\left|\psi_{1}\right \rangle$ and $\left|\psi_{2}\right\rangle$ are the standard Bell states that remain unchanged with time. We will show in this work that the former ones play a fundamental role in the behaviour of the measurement-based cycle. Note that in the limit of $\gamma =0$, the eigenstates $|\psi_{0,3}\rangle$ take the form of product (i.e., disentangled) states, with the respective eigenvalues $\mp 2B$. [see \textbf{App.~\ref{Eigenstate isotropic}}].

\begin{figure}[h!]
 \includegraphics[width=0.60\textwidth]{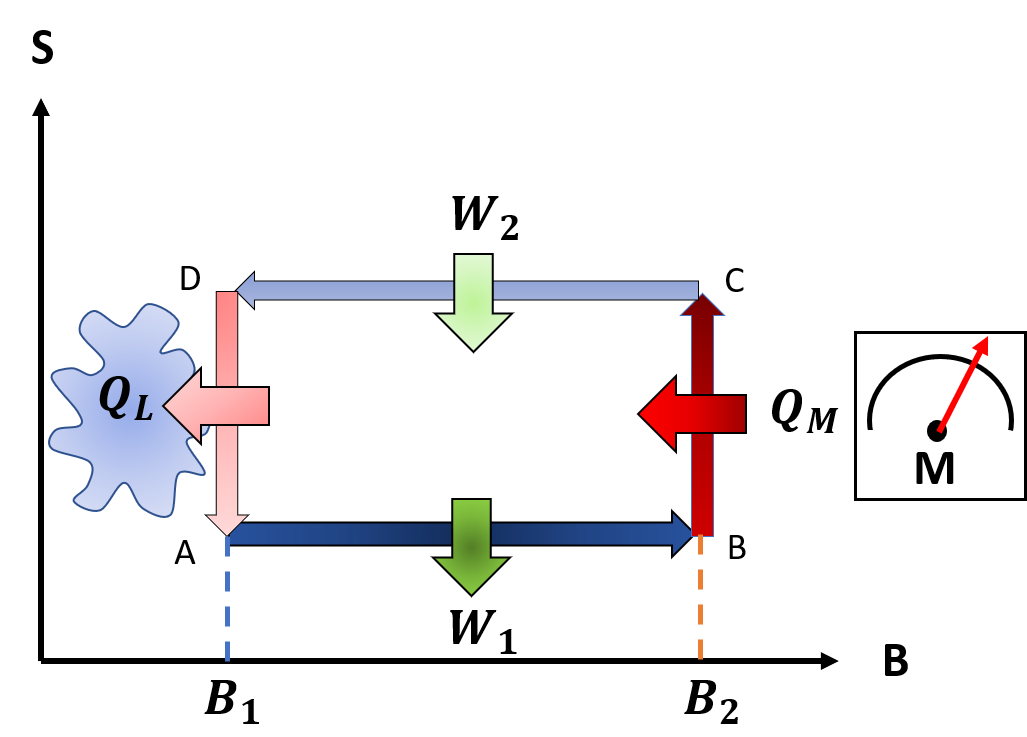}
   \caption{Schematic diagram of the Otto cycle}
   \label{fig:schematic diagram}
\end{figure}

\subsection{Quantum Otto cycle and thermodynamic quantities}\label{cycle}

We consider that the working system undergoes an Otto cycle. The schematic diagram of the cycle is shown in the \textbf{Fig.~\ref{fig:schematic diagram}}. The strokes of the cycle are described below.

{\bf Unitary expansion (A to B)}: The working system is initially prepared in a thermal state $\hat{\rho}_{A}=e^{-\beta \hat{H}_{1}} / Z$ at inverse temperature $\beta = 1/T\left(k_{B}=1\right)$, with $\hat{H}_{1}=\hat{H}(0)$ and $Z=\operatorname{Tr}(e^{-\beta \hat{H}_{1}})$. During this stage of the cycle, the system is decoupled from the heat bath and the external magnetic field is changed from $B_{1}$ to $B_{2}$ during a finite time-interval $\tau$. We choose a linear ramp for this change:  $B(t) = B_{1} + (B_{2} - B_{1})(t/\tau)$, where $0 \leqslant t \leqslant \tau$. 
The state of the working system at the end of this stage changes to $\hat{\rho}_{B}=\hat{U}(\tau) \hat{\rho}_{A} \hat{U}^{\dagger}(\tau)$, where $\hat{U}(\tau)=\mathcal{T} \exp[-\iota \int_{0}^{\tau} d t \hat{H}(t)]$ is the relevant time evolution operator, with $\mathcal{T}$ indicating the time-ordering.  
Also, a certain amount of work, $W_{1}$ is done by the system, which can be calculated as $W_{1}=\langle E_{B}\rangle - \langle E_{A}\rangle$, where $\langle E_{A}\rangle=\operatorname{Tr}(\hat{\rho}_{A} \hat{H}_{1})$ and  $\langle E_{B}\rangle=\operatorname{Tr}(\hat{\rho}_{B} \hat{H}_{2})$ indicate the expectation values of the internal energies of the system at the start and the end of this stage. Note that $\hat{H}_{2}=\hat{H}(\tau)$.

{\bf Isochoric heating (B to C)}: The heating of a system can be generally understood to be associated with an increase in its entropy. 
Usually, a system is heated using a heat bath. This can be alternatively achieved by applying a non-selective quantum measurement on the working system. In order to ensure that the energy supplied by this measurement is nonzero, the measurement operator $\hat{M}$ should not commute with the Hamiltonian, i.e. $[\hat{H}\left(B_{2}\right), \hat{M}] \neq 0$. If $\hat{\rho}$ is the state before the measurement, the post-measurement state is usually written as $\sum_{\alpha} \hat{M}_{\alpha} \hat{\rho} \hat{M}_{\alpha}$, where $\hat{M}_{\alpha}=|M_{\alpha}\rangle\langle M_{\alpha}|$ is the projection operator associated to the non-degenerate eigenvalues of the observable $M$ with eigenstates $|M_{\alpha}\rangle$, satisfying $\hat{M}_{\alpha}^{\dagger}=\hat{M}_{\alpha}$ and $\sum_{\alpha} \hat{M}_{\alpha}^{2}=\mathds{1}$. 
In the present case, we perform a global measurement of the state of the system \cite{huangQIP2020}, in the Bell basis $\{\left|\psi_\pm\right\rangle=\frac{1}{\sqrt{2}}(|00\rangle \pm|11\rangle)$,
$\left|\phi_\pm\right\rangle=\frac{1}{\sqrt{2}}(|01\rangle \pm|10\rangle)\}$. The leads to a post-measurement state given by $\hat{\rho}_{C} =\sum_{\alpha = 1}^4 \hat{M}_{\alpha} \hat{\rho}_{B} \hat{M}_{\alpha}$. where $\hat{M}_\alpha$ describes the relevant projection operators as follows: $\hat{M}_{1,2}=\left|\psi_\pm\right\rangle\left\langle\psi_\pm\right|$ and $\hat{M}_{3,4}=\left|\phi_\pm\right\rangle\left\langle\phi_\pm\right|$.
During this stage, the entropy of the system increases due to its interaction with the measurement apparatus and this increase can be considered equivalent to heating. The corresponding heat `absorbed' can be calculated as $Q_{M} = \langle E_{C}\rangle - \langle E_{B}\rangle$, where the internal energy $\langle E_{C}\rangle = \operatorname{Tr}(\hat{\rho}_{C}\hat{H}_{2})$. 


{\bf Unitary compression (C to D)}: The working system remains decoupled from the heat bath in this stage. The magnetic field is driven from $B_{2}$ to $B_{1}$ in a finite time $\tau$ using the protocol $B(\tau - t)$. The state of the working system at the end of this stage becomes $\hat{\rho}_{D}=\hat{V}(\tau) \hat{\rho}_{C} \hat{V}^{\dagger}(\tau)$, where $\hat{V}(\tau)=\mathcal{T} \exp[-\iota \int_{0}^{\tau} d t \hat{H}(\tau-t)]$ is the time evolution operator. 
A certain amount of work, $W_{2}$, is done on the system, which can be calculated as $W_{2}=\langle E_{D}\rangle - \langle E_{C}\rangle$, where the internal energy $\langle E_{D}\rangle=\operatorname{Tr}(\hat{\rho}_{D} \hat{H}_{1})$.


{\bf Isochoric cooling (D to A)}: During this final stage of the cycle, the system is now coupled with a heat bath at the temperature $T$, whereas the magnetic field remains fixed at $B_{1}$. The system releases some amount of heat $Q_{L}$ to the bath, which can be calculated as $ Q_{L} = \langle E_{A}\rangle - \langle E_{D}\rangle$.
We assume that this process is carried out over a long time so that the system reaches thermal equilibrium with the bath.


Total work done in a complete cycle can be calculated as $W = (W_{1} + W_{2}) = -(Q_{M} + Q_{L})$. If $W < 0$, then the total work in a complete cycle is done by the working system. Also, the working system absorbs some amount of heat in the measurement process, if $Q_{M} > 0$. Then, the working system in a complete cycle works as a heat engine. So, the efficiency of the engine is given by $\eta =  |W|/Q_{M}$.

\section{Finite time operation of the engine}\label{finite time operation}
\begin{figure}[h!]
 \includegraphics[width=0.50\textwidth]{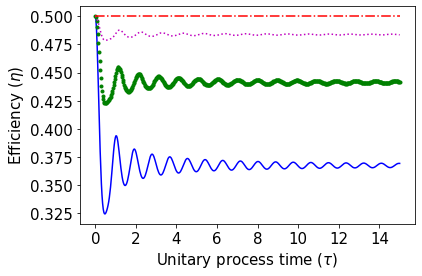}
   \caption{Efficiency as a function of duration $\tau$ of the unitary stages, for different values of the anisotropy parameter $\gamma = 0$ (dash-dotted red line), $\gamma = 0.3$ (point-marked magenta line), $\gamma = 0.6$ (dotted green line), $\gamma = 1$ (solid blue line). The other parameters are $B_{1} = 1$, $B_{2} = 2$, $T = 1$. All the quantities are dimensionless with respect to $J$ and also we have used $k_{B} = \hbar = 1$.}
   \label{fig:Eff Vs Unitary time 2}
\end{figure}

Usually, quantum heat engines are studied quasistatically. If we allow different stages of the engine cycle only for finite times, the performance of the engine is expected to deviate substantially from the steady state. We show in \textbf{Fig.~\ref{fig:Eff Vs Unitary time 2}}, how the efficiency varies with respect to the duration of the unitary stages. We assume that each of these stages (unitary expansion and compression) occurs for the same duration $\tau$. All simulations are done using QuTip \cite{johansson2012CPC} software package.

As seen in the \textbf{Fig.~\ref{fig:Eff Vs Unitary time 2}}, the  efficiency oscillates at the transient time-scale  for $\gamma \neq 0$. This means that if the unitary stages are executed for a very short time $\tau \gtrsim 0$, the efficiency can be larger or smaller than that obtained for a large value of $\tau$. 
If the unitary processes are prolonged, the oscillation in efficiency disappears. Thus, a finite-time measurement-based engine can perform better than the same engine operating for a longer duration, for a suitable selection of the duration $\tau$ for unitary processes.

Note that, if one would use a local measurement, instead of global ones, similar oscillatory behaviour in the efficiency of the engine could be seen, for finite-time operation \cite{das2019Entropy,huang2020QIP}. Also in these cases, the engine performs better than its quasistatic counterpart at finite times for specific choices of the local basis.

\subsection{Thermodynamic quantities in terms of transition probabilities}\label{tran prob}
The results as mentioned above can further be analyzed in terms of the transition probabilities between the instantaneous eigenstates of the Hamiltonian. The internal energies (derivations are given in the \textbf{App~\ref{ApeA}}) of the system at four vertices of the QHE diagram in \textbf{Fig.~\ref{fig:schematic diagram}}, are given by
\begin{equation}\label{int ene}
\begin{aligned}
&\langle E_{A}\rangle=-4 K_{1} \frac{\sinh 2 K_{1} \beta}{Z}-4 J \frac{\sinh 2J \beta}{Z}\;,\\
&\langle E_{B}\rangle=- 4K_{2} (1 - 2\xi) \frac{ \sinh 2 K_{1} \beta}{Z}-4J \frac{\sinh 2J \beta}{Z}\;, \\
&\langle E_{C}\rangle=-4 K_{2} (1 - 2\delta) (1 - 2\chi) \frac{ \sinh 2 K_{1} \beta}{Z}-4J \frac{\sinh 2J \beta}{Z}\;,\\
&\langle E_{D}\rangle=-4 K_{1} (1 - 2\delta) (1 - 2\lambda) \frac{ \sinh 2 K_{1} \beta}{Z}-4J \frac{\sinh 2J \beta}{Z}\;,
\end{aligned}
\end{equation}
where $Z = 2\cosh(2K_{1}\beta) + 2\cosh(2J\beta)$ is the partition function, $K_{1}=\sqrt{B_{1}^{2}+\gamma^{2}J^{2}}$, $K_{2}=\sqrt{B_{2}^{2}+\gamma^{2}J^{2}}$, $\xi=|\langle\psi_{0}^{(2)}|\hat{U}(\tau)| \psi_{3}^{(1)}\rangle|^{2}$,  $\delta=|\langle\psi_{+}|\hat{U}(\tau)| \psi_{0}^{(1)}\rangle|^{2}$, $
\chi=|\langle\psi_0^{(2)}| \psi_{+}\rangle|^2
$, and
$\lambda=|\langle\psi_{3}^{(1)}|\hat{V}(\tau)| \psi_{-}\rangle|^{2}$. Clearly, $\xi$ accounts for the transition probability between two different eigenstates $|\psi_3\rangle$ and $|\psi_0\rangle$ during the unitary expansion. Also, because the instantaneous energy eigenstates $|\psi_{0,3}\rangle$ do not truly coincide with the measurement basis states $|\psi_\pm\rangle$, their nonzero overlap gives rise to certain transition between them during 
measurement and unitary compression stages of the cycle. This can be seen by rewriting the states $|\psi_\pm\rangle$ in terms of the instantaneous energy eigenstates, as
\begin{equation}
\begin{aligned}
&|\psi_+\rangle =  - \frac{c_2-d_2}{a_2d_2-b_2c_2}|\psi_{0}^{(2)}\rangle + \frac{a_2-b_2}{a_2d_2-b_2c_2} |\psi_{3}^{(2)}\rangle\\
&|\psi_-\rangle =  - \frac{c_2+d_2}{a_2d_2-b_2c_2} |\psi_{0}^{(2)}\rangle + \frac{a_2+b_2}{a_2d_2-b_2c_2} |\psi_{3}^{(2)}\rangle,
\end{aligned}
\end{equation}
where
\begin{eqnarray}\label{abcd}
a_2 = \frac{B_{2} - K_{2}}{\sqrt{K_{2}^{2}  - B_{2}K_{2}}}, b_2 = \frac{\gamma J}{\sqrt{K_{2}^{2} - B_{2}K_{2}}}, c_2 = \frac{B_{2} + K_{2}}{\sqrt{K_{2}^{2} + B_{2}K_{2}}}, d_2 = \frac{\gamma J}{\sqrt{K_{2}^{2} + B_{2}K_{2}}}\;.
\end{eqnarray}
Then the relevant transition probabilities can be written as
\begin{equation}\label{delta labda}
\begin{aligned}
&\delta=\left|-\frac{c_2-d_2}{a_2 d_2-b_2 c_2}\langle\psi_0^{(2)}|U(\tau)| \psi_0^{(1)}\rangle +\frac{a_2-b_2}{a_2 d_2-b_2 c_2}~\langle\psi_3^{(2)}|U(\tau)| \psi_0^{(1)}\rangle\right|^2 \;,\\
&\lambda=\left|-\frac{c_2+d_2}{a_2 d_2-b_2 c_2}\langle\psi_3^{(1)}|V(\tau)| \psi_0^{(2)}\rangle+\frac{a_2+b_2}{a_2 d_2-b_2 c_2}\langle\psi_3^{(1)}|V(\tau)| \psi_3^{(2)}\rangle\right|^2\;,
\end{aligned}   
\end{equation}
and
\begin{equation}\label{prob chi}
   \chi=  \left|\frac{c_2-d_2}{a_2d_2-b_2c_2}\right|^2. 
\end{equation}

The expressions of the work can be obtained using the  \textbf{Eq.~\ref{int ene}} as $$W_{1}= 4[ K_{1}-K_{2} (1 - 2\xi)] \frac{ \sinh 2 K_{1} \beta}{Z} $$ and $$W_{2}=4 (K_2-K_{1}) (1 - 2\delta) (1 - 2\lambda) \frac{ \sinh 2 K_{1} \beta}{Z}.$$ Thus, the total work in a complete cycle is given by 
\begin{equation}\label{work}
  W = W_1 + W_2 = -4[K_{2}\{(1 - 2\xi) - (1 - 2\delta)(1 - 2\chi)\} - K_{1}\{1 - (1 - 2\delta)(1 - 2\lambda)\}] \frac{ \sinh 2 K_{1} \beta}{Z}.
\end{equation}
Also, the heat `absorption' during the measurement stroke is given by 
\begin{equation}\label{heat}
  Q_{M} = 4K_{2}[(1 - 2\xi) - (1 - 2\delta)(1 - 2\chi)]\frac{ \sinh 2 K_{1} \beta}{Z}.
\end{equation}
The efficiency of the cycle is therefore given by 
\begin{equation}\label{eff}
  \eta =  \frac{|W|}{Q_{M}}= 1 - \frac{K_{1}[1 - (1 - 2\delta)(1 - 2\lambda)]}{K_{2}[(1 - 2\xi) - (1 - 2\delta)(1 - 2\chi)]}\;.  
\end{equation}

\begin{figure}[h!]
\includegraphics[width=0.50\textwidth]{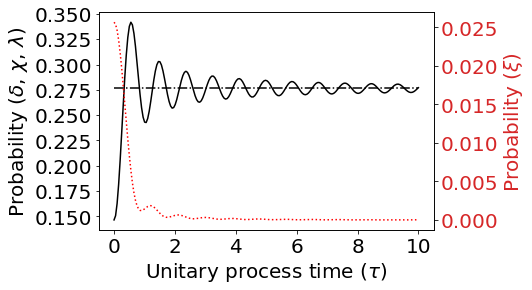}   
\caption{Transition probabilities as a function of the duration $\tau$ of each unitary stage. We have used the left y-axis for transition probabilities $\delta$ and $\lambda$  (solid black line), and $\chi$ (dash-dotted black line) and the right y-axis for transition probability $\xi$ (point-marked red line). The other parameters are $B_{1} = 1$, $B_{2} = 2$, $\gamma = 1$.}
   \label{fig:Transition prob Vs Unitary time}
\end{figure}

The plot of the transition probabilities $\xi$, $\delta$, $\chi$, and $\lambda$ with respect to the duration $\tau$ of the individual unitary processes are shown in the \textbf{Fig.~\ref{fig:Transition prob Vs Unitary time}}. Note that $\delta$ and $\lambda$ exhibit oscillatory dependence on $\tau$, $\chi$ remains constant, while $\xi$ displays a monotonic decay, as $\tau$ increases. Though the transition probabilities $\delta$ and $\lambda$ are the same, for our choice of the measurement basis and the eigenstates of the Hamiltonian, it is, generally speaking, not a universal feature \cite{lin2021PRA}. 

The oscillation in the finite time efficiency is primarily due to the oscillation in the transition probabilities $\delta$ and $\lambda$. The oscillation in the transition probabilities $\delta$ can be attributed to the interference between the probability amplitudes for the transitions $|\psi_0^{(1)}\rangle \rightarrow |\psi_0^{(2)}\rangle$ and $|\psi_0^{(1)}\rangle \rightarrow |\psi_3^{(2)}\rangle$  (see \textbf{Eq.~\ref{delta labda}}). Similarly, oscillation in $\lambda$ is due to the transition $|\psi_-\rangle \leftrightarrow |\psi_3^{(1)}\rangle$. The other transition probability $\xi$, also  known as quantum internal friction \cite{jiao2021PRE,rezek2010Ent}, is negligible with respect to the $\delta$, $\chi$ and $\lambda$ in a measurement-based QHE. As the unitary stages are prolonged, oscillation in the finite time efficiency disappears and the efficiency approaches to the quasistatic limit (see \textbf{Sec.~\ref{adiabatic}}). 

Note that the two energy eigenstates $\left|\psi_{1}\right\rangle$ and $\left|\psi_{2}\right\rangle$ are the same as two Bell states  $\left|\phi_\pm\right\rangle$. As these states have been used in measurement basis, their occupation probabilities do not change in the measurement process, and therefore, these states do not contribute to the calculation of heat [see Eq. (\ref{delta labda}), (\ref{prob chi}), and (\ref{heat})]. Moreover, the eigenvalues of these eigenstates are independent of the external control parameter $B(t)$, and thus the contribution of these states to work is also zero [see Eq. (\ref{work})]. The only contribution to the engine performance arises from the two other eigenstates $\left|\psi_{0}\right\rangle$ and $\left|\psi_{3}\right\rangle$.

\subsection{Quasistatic (adiabatic) limit of the thermodynamic quantities}\label{adiabatic}

In order to calculate the quasistatic value of the efficiency, we consider that the unitary processes are performed quasistatically, i.e., for an infinite time interval. Therefore, there are no nonadiabatic transitions between two instantaneous energy eigenstates, and the unitary processes become adiabatic. So, in such limit, we can write,  $\xi=|\langle\psi_{0}^{(2)}|\hat{U}(\tau)| \psi_{3}^{(1)}\rangle|^{2} \stackrel{\tau \rightarrow \infty}{=} 0 $  (\textbf{Fig.~\ref{fig:Transition prob Vs Unitary time}}). Also, the transition probabilities between the instantaneous energy eigenstates and the basis states of measurement take the following forms for very large $\tau$:
$$\delta \stackrel{\tau \rightarrow \infty}{=} \left|\frac{c_2-d_2}{a_2d_2-b_2c_2}\right|^{2}\;\;,\;\;\;   \lambda \stackrel{\tau \rightarrow \infty}{=} \left|  \frac{a_2+b_2}{a_2d_2-b_2c_2}\right|^{2}.$$
Using the expressions of $a_2,b_2,c_2,d_2$ (see \textbf{Eq.~\ref{abcd}}) we can indeed find that $\delta= \chi = \lambda = \frac{1}{2} -\gamma J /2K_{2}$, at  the quasistatic limit (see also \textbf{Fig.~\ref{fig:Transition prob Vs Unitary time}}). 

Thus, the expressions of the work and heat absorption of the cycle can be obtained as 
\begin{equation}
\begin{aligned}
&W_{q} = -16(K_{2} - K_{1})\chi(1 - \chi) \frac{ \sinh 2 K_{1} \beta}{Z}, \\
&Q_{Mq} = 16K_{2}\chi (1 - \chi)\frac{ \sinh 2 K_{1} \beta}{Z}\;,
\end{aligned}
\end{equation}
and the quasistatic value of the efficiency is given by 
\begin{equation}\label{effq}
  \eta_{q} =|W_q|/Q_{Mq} = 1 - \frac{K_{1}}{K_{2}}\;.
\end{equation}

Clearly, the expression of this efficiency is independent of the temperature of the heat bath used in the cold isochoric process. This indicates that the performance of the engine does not depend upon the temperature of the heat bath in the case of global measurement, which we have used in the isochoric heating stage. We emphasize that this is unlike the case for a local measurement where the performance of an engine depends upon the temperature of the heat bath \cite{das2019Entropy}. Also, it can be seen from the \textbf{Eq.~\ref{effq}} that for nonzero $\gamma$, the expression of the efficiency is very much similar to the efficiency of a single-spin QHE with two heat baths \cite{dann2020Entropy} or a single heat bath and a non-selective quantum measurement at the isochoric heating stage \cite{yiPRE2017}.
This similarity arises as only two intermediate energy levels ($|\psi_{0,3}\rangle$, as mentioned in \textbf{Sec.~\ref{model}}) contribute to the engine performance, due to our specific choice of the measurement basis. Therefore, a measurement-based heat engine with a coupled two-spin working system for global measurement acts like a two-level (single-spin) heat engine, which is evident in the expression of the efficiency (\textbf{Eq.~\ref{effq}}). Interestingly, even a two-stroke QHE made up of two different working systems with two different frequencies can lead to the same form of efficiency \cite{piccione2021PRA}. However, the expression of efficiency will differ if one uses a coupled two-spin working system along with two heat baths or with a single bath plus local measurement instead of global measurement.

\subsection{Sudden (quench) limit of the thermodynamic quantities}\label{sudden}
In order to calculate the sudden limit of the thermodynamic quantities, we consider that the external magnetic field is changed suddenly ($\tau \rightarrow 0$) from $B_1$ to $B_2$ or vice versa. In this case $\hat{U}(\tau), \hat{V}(\tau) \rightarrow \mathds{1}$, therefore the state of the system does not change over unitary processes. So, in this limit, the transition probabilities can be written as
\begin{eqnarray}
\delta & \stackrel{\tau \rightarrow 0}{=}&\left|-\frac{c_2-d_2}{a_2 d_2-b_2 c_2}{ }~\langle\psi_0^{(2)}|\psi_0^{(1)}\rangle+\frac{a_2-b_2}{a_2 d_2-b_2 c_2}~\langle\psi_3^{(2)}|\psi_0^{(1)}\rangle\right|^2 = - \frac{(B_1 - K_1 + \gamma J)^2}{4K_1(B_1 - K_1)},\nonumber\\
\lambda &\stackrel{\tau \rightarrow 0}{=}&\left|-\frac{c_2+d_2}{a_2 d_2-b_2 c_2}~\langle\psi_3^{(1)}| \psi_0^{(2)}\rangle+\frac{a_2+b_2}{a_2 d_2-b_2 c_2}~\langle\psi_3^{(1)}| \psi_3^{(2)}\rangle\right|^2 
=   \frac{(B_1 + K_1 - \gamma J)^2}{4K_1(B_1 + K_1)},\label{dellamxi}\\
\xi &\stackrel{\tau \rightarrow 0}{=}& \left|\langle\psi_{0}^{(2)}|\psi_{3}^{(1)}\rangle\right|^{2} 
= - \frac{\gamma^2 J^2 + (B_1+K_1)(B_2 - K_2)}{4K_1K_2 (B_2 - B_1)(B_1 + K_1)}\;.\nonumber
\end{eqnarray}
Also from the \textbf{Eq.~\ref{prob chi}}, we get
\begin{equation}
\chi= \frac{(B_2 - K_2)(B_2 + K_2 - \gamma J)^2}{4 \gamma J^2 K_2}\;.\label{chi}
\end{equation}

Using (\ref{dellamxi}) and (\ref{chi}) in the expressions of work (\textbf{Eq.~\ref{work}}) and heat absorption (\textbf{Eq.~\ref{heat}}), we obtain the sudden limits of work and heat absorption which are given by
\begin{equation}
\begin{aligned}
&W_{s} =  -4\frac{B_1(B_2 - B_1)}{K_1} \frac{ \sinh 2 K_{1} \beta}{Z}, \\
&Q_{Ms} = 4\frac{B_1 B_2}{K_1}\frac{ \sinh 2 K_{1} \beta}{Z}\;.
\end{aligned}
\end{equation}
Therefore, the efficiency at this limit is given by
\begin{equation}
    \eta_s = \frac{|W_s|}{Q_{Ms}}= 1 - \frac{B_1}{B_2}.
\end{equation}

Interestingly, the efficiency does not depend on the anisotropy parameter in the sudden limit. This means that the QOE has the same efficiency for all $\gamma$ when $\tau\rightarrow 0$. 
This can be seen in the \textbf{Fig.~\ref{fig:Eff Vs Unitary time 2}} that for our choices of $B_1 =1$ and $B_2 = 2$, this is equal to 0.5, irrespective of the values of $\gamma$, whereas for large $\tau$, the efficiency saturates to a lower value.  Therefore, with two spins coupled by anisotropic interaction as the working system, a measurement-based QHE operating in the sudden limit performs better than an engine operating in the adiabatic limit.


\subsection{Analysis of the heat engine performance}\label{analysis}

\begin{figure}[h!]
\includegraphics[width=0.50\textwidth]{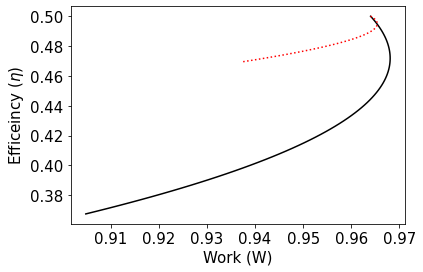}
   \caption{The parametric plot of the variable $\gamma$ on the work-efficiency plane. We have taken the absolute value of the work. Here the point-marked red line represents the finite-time value (for $\tau = 0.1 $) and the solid black line represents the quasistatic value. The anisotropy parameter $\gamma$ varies from $0$ to $1$. The point $0.5$ on the solid black line corresponds to $\gamma = 0$, while the left end of the plot corresponds to $\gamma = 1$. The other parameters are $B_{1} = 1$, $B_{2} = 2$, and $T = 1$.}
   \label{fig:Work heat eff Vs anisotropy paramteter 1}
\end{figure}

For the quasistatic operation of the cycle, we show the variation of the efficiency as a function of work in the \textbf{Fig.~\ref{fig:Work heat eff Vs anisotropy paramteter 1}}. It is clear from this plot that the engine performance degrades with the increase of the anisotropy parameter $\gamma$. 
This is because, as $\gamma$ increases, the heat absorption in the measurement process increases and the work output decreases after a slow increase in the lower range of $\gamma$. We also observed that there exists a certain value of $\gamma \sim 0.46$, for which the work output gets maximized.

\begin{figure}[h!]
\includegraphics[width=0.50\textwidth]{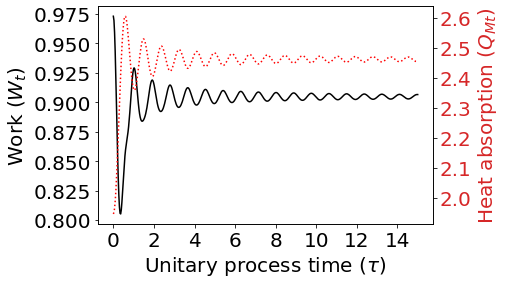}
   \caption{Variation of the absolute value of the work done $W_{t}$ (solid black line, labelled on the left y-axis) and heat absorbed $Q_{Mt}$ (point-marked red line, labelled on the right y-axis) as a function of duration $\tau$ of the unitary stage.  The other parameters are $B_{1} = 1$, $B_{2} = 2$, $T = 1$, and $\gamma = 1 $. }
   \label{fig:Work and heat Vs Unitary time}
\end{figure}

The variation of the work and heat absorption with respect to the duration $\tau$ of the unitary processes are shown in the \textbf{Fig.~\ref{fig:Work and heat Vs Unitary time}}. 
Also, the plots of the efficiency with respect to work are shown in the \textbf{Fig.~\ref{fig:Work heat eff Vs anisotropy paramteter 1}}. From these plots, we can see that a finite-time engine can deliver more work than the same engine operating in the quasistatic limit with a proper choice of the time interval $\tau$ of the unitary processes. In addition to that, the finite-time engine absorbs less amount of heat in the measurement process than the same engine operating in the quasistatic limit. Consequently, when operated for finite times, the engine requires less energy resource, and still can perform better than its quasistatic counterpart.

Interestingly, this outperforming is further improved for larger anisotropy parameter $\gamma$. When $\gamma = 0$, no transition takes place between two instantaneous energy eigenstates, and the unitary stages remain adiabatic, irrespective of their duration. Also, there is no interference-like effect between two transition probability amplitudes as for the anisotropic case, which can be seen in \textbf{App.~\ref{Eigenstate isotropic}}. Thus, the efficiency does not change with respect to $\tau$, as displayed in the \textbf{Fig.~\ref{fig:Eff Vs Unitary time 2}}. Therefore, operating the engine even for a finite time would lead to the same efficiency as for the case when operated quasistatically.

\section{Always-on coupling to the heat bath}\label{all time bath on}

It may not always be possible to decouple a quantum system from its bath, which acts as a heat bath for the HE operation, depending upon the architecture of the working system and the bath. Also, there is a cost associated with coupling and decoupling the working system from a heat bath \cite{kurizki2022book, bhattacharjee2021EPJB}. In the previous section, we assumed that the working system is completely isolated from its bath during the work-delivering stages, so that the stages AB and CD remain unitary. We consider here that the HE operation is implemented in a type of realistic architecture in which the working system cannot be decoupled from its bath \cite{kurizki2022book}. It is therefore necessary to take into account the dissipation of energy from the working system to the bath during the stages AB and CD. This requires solving the master equation, which is given below, with a time-dependent Hamiltonian under a dissipative bath. Here, we assume that the remaining two isochoric stages of the cycle are identical to those mentioned in the \textbf{Sec.~\ref{cycle}}. Furthermore, since the measurement process is assumed to be instantaneous, the bath will not have any effect on the system during measurement.

We consider that the temperature of the heat bath is $T$ and a single spin decay to the bath. Then the master equation in the interaction picture for two spins can therefore be written as \cite{huang2012PRE} 
\begin{equation}
\begin{aligned}
\frac{\partial \hat{\rho}}{\partial t} =& \iota[\hat{\rho},\hat{H}] + \sum_{i=1,2}\left[\Gamma_i(t)\left\{n(\omega_i (t))+1\right\}\left(\hat{X}_i \hat{\rho} \hat{X}_i^{\dag}-\frac{1}{2} \hat{X}_i^{\dag} \hat{X}_i \hat{\rho}-\frac{1}{2} \hat{\rho} \hat{X}_i^{\dag} \hat{X}_i\right)\right.\\
&\left.+~\Gamma_{i}(t) n\left(\omega_i (t)\right)\left(\hat{X}_i^{\dag} \hat{\rho} \hat{X}_i-\frac{1}{2} \hat{X}_i \hat{X}_i^{\dag} \hat{\rho}-\frac{1}{2} \hat{\rho} \hat{X}_i \hat{X}_i^{\dag}\right)\right]\;,
\end{aligned}   
\label{master}
\end{equation}
where $\Gamma_i(t)\left\{n\left(\omega_i (t)\right)+1\right\}$ and $\Gamma_i(t)n\left(\omega_i (t)\right)$ are time-dependent dissipation rates, $n\left(\omega_i (t)\right)=\left[\exp \left(\frac{\hbar \omega_i (t)}{k T}\right)-1\right]^{-1}$ is the average number of photons in the bath at the transition frequencies $\omega_i (t)$, $\Gamma_i(t)$ = $0.1 \omega_i(t) e^{-\omega_i(t)/\omega_c}$ is the time-dependent dissipation rate of the spin to an Ohmic type bath \cite{dann2018PRA,leggett1987RMP}, and $\omega_c$ is the cut-off frequency of the bath spectral density. These rates are time-dependent because of the time-dependence of the Hamiltonian $\hat{H}(t)$ \cite{dann2018PRA,albash2012NJP}. 
Note that we are assuming that the \textbf{Eq.~\ref{master}} remains valid for the time scales involved with the system and the bath dynamics. This is possible if the bath time scale $1/\omega_c$ is much smaller than the system time-scale $1/{\rm min}_j (E_j)$ [where $E_j$ is given by \textbf{Eq.~\ref{eigen}}] and the duration $\tau$ during which the magnetic field is changed \cite{dann2018PRA}.

Note that the jump operators associated to a system operator $\hat{X}$ are given by \cite{breuer2002book}  
$$
\hat{X}(\omega)=\sum_{\epsilon^{\prime}-\epsilon=\omega}|\epsilon\rangle\langle\epsilon|\hat{X}| \epsilon^{\prime}\rangle\langle\epsilon^{\prime}|,
$$
where $\{|\epsilon\rangle\}$ is the basis of the eigenvectors of the system Hamiltonian $\hat{H}$. In the present case, the first spin is assumed to interact with the heat bath via $\hat{\sigma}^{x}$ operator, and the corresponding jump operators $\hat{X}_i$ and the respective transition frequencies are given by
\begin{equation}\label{jump}
\begin{array}{ll}
\hat{X}_1=\frac{1}{2}\left( \frac{B+K-\gamma J}{\sqrt{K^2 + BK}}~ \left|\psi_1\right\rangle\left\langle\psi_3\right| + \frac{B-K+\gamma J}{\sqrt{K^2 - BK}}~ \left|\psi_0\right\rangle\left\langle \psi_2\right|\right), \quad \hbar \omega_1 = 2K + 2J \\
\hat{X}_2=\frac{1}{2}\left( \frac{B+K+\gamma J}{\sqrt{K^2 + BK}}~ \left|\psi_2\right\rangle\left\langle\psi_3\right| + \frac{B-K-\gamma J}{\sqrt{K^2 - BK}}~ \left|\psi_0\right\rangle\left\langle \psi_1\right|\right), \quad \hbar \omega_2= 2K - 2J.
\end{array}
\end{equation}
These operators satisfy the relations $[\hat{H}, \hat{X}_i]= - \omega_i \hat{X}_i$ and $[\hat{H}, \hat{X}_i^{\dag}]= \omega_i \hat{X}_i^{\dag}$. Note that $K$ and $|\psi_{0,3}\rangle$ are functions of $B(t)$, and are therefore time-dependent, which also gives rise to the time dependence of the jump operators in \textbf{Eq.~\ref{jump}}.

The heat and work in an open quantum system in presence of an external derive are defined as \cite{kurizki2022book}
\begin{equation}\label{finite time heat and work}
\begin{array}{ll}
Q(t)=\int_0^t \operatorname{Tr}\left[\dot{\hat{\rho}}\left(t^{\prime}\right) \hat{H}\left(t^{\prime}\right)\right] d t^{\prime}\;,\\
W(t)=\int_0^t \operatorname{Tr}\left[\hat{\rho}\left(t^{\prime}\right) \dot{\hat{H}}\left(t^{\prime}\right)\right] d t^{\prime}.
\end{array}
\end{equation}
The total change in the average energy of the system in a process is given by $ \Delta E(t)=E(t)-E(0)$, where $E(t)=\operatorname{Tr}[\hat{\rho}(t) \hat{H}(t)]$ is the average energy at a time $t$. This change in energy can be written as contributions from two separate thermodynamic quantities as 
\begin{equation}\label{finite time energy}
  \Delta E(t)=W(t)+Q(t).  
\end{equation}

\begin{figure}[h!]
\includegraphics[width=0.50\textwidth]{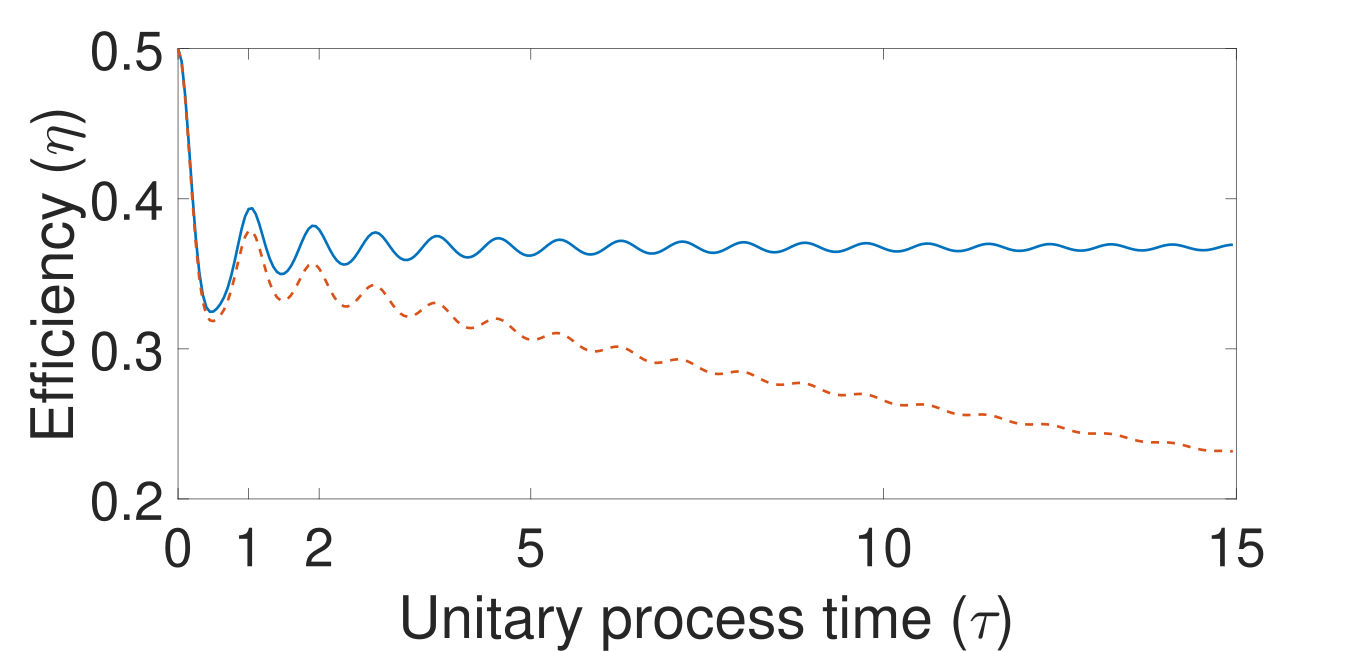}
   \caption{Efficiency $\eta$ as a function of duration $\tau$ of the unitary stages for $\Gamma_i(t) = 0$
   (for the isolated system: solid blue line) and 
   $\Gamma_i(t)$ = $0.1 \omega_i(t) e^{-\omega_i(t)/\omega_c}$ (for open system: dashed red line). The other parameters are $B_{1} = 1$, $B_{2} = 2$, $T = 1$, $\omega_c = 1000$, and $\gamma = 1$.} 
   \label{fig:Work heat and efficiency Vs Unitary time}
\end{figure}

The work in the AB and CD processes can then be represented as follows.
\begin{equation}
\begin{array}{ll}
W_{1}(\tau)=\int_0^{\tau} \operatorname{Tr}\left[\hat{\rho}_{A\rightarrow B}\left(t^{\prime}\right) \dot{\hat{H}}_{A\rightarrow B}\left(t^{\prime}\right)\right] d t^{\prime}\\
W_{2}(\tau)=\int_0^{\tau} \operatorname{Tr}\left[\hat{\rho}_{C\rightarrow D}\left(t^{\prime}\right) \dot{\hat{H}}_{C\rightarrow D}\left(t^{\prime}\right)\right] d t^{\prime}.
\end{array}
\end{equation}
So, the total work is given by $W = W_{1} + W_{2}$. We calculated the heat transfer between the system and the heat bath using the 
\textbf{Eq.~\ref{finite time energy}}. The numerical solution of the master equation has been done using the 4th-order adaptive Runge-Kutta method and the numerical integration to calculate the work is done using the Trapezoidal rule. The heat absorption in the measurement process is calculated as discussed in the \textbf{Sec.~\ref{cycle}}.

We show in the \textbf{Fig.~\ref{fig:Work heat and efficiency Vs Unitary time}} how the efficiency varies with the duration $\tau$ of the unitary stages. It is clear from this plot that the presence of the bath has a negligible effect on its performance in a very short time. We can, therefore, employ such a measurement-based heat engine model whenever one requires a finite amount of power. One does not have to decouple the working system ever to obtain a finite amount of power, if the engine runs for a finite duration. However, the longer the duration $\tau$, the engine efficiency decreases, due to the dominant effects of the bath over the external control parameter. The dissipative part of the master equation dominates over the unitary part and therefore, the system releases more energy to the bath as heat than it releases as the work. Also, the spins absorb more energy during the measurement stage as $\tau$ increases. We must emphasize that if both the spins are considered to individually interact with the heat bath \cite{hu2018QIP,liao2011PRA}, the main results of this paper will remain the same.

In this section, we have considered the QOE with an always-on single bath along with a measurement protocol.  
On the other hand, it is possible for a QOE that operates with two heat baths to maintain such an always-on coupling, while still achieving a reciprocating cycle by periodically changing the interaction strength with the baths. However, as the performance of these QOEs deteriorates in a finite time, such an always-on interaction cannot give us operational advantages over a measurement-based engine.

\subsection{Power analysis of the engine}

\begin{figure}[h!]
 \includegraphics[width=0.50\textwidth]{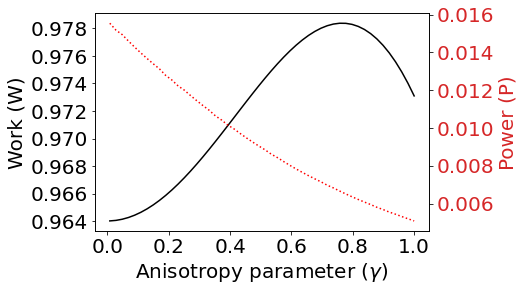}
   \caption{Variation of the absolute value of work (black solid line, labelled on the left y-axis) and power (point-marked red line, labelled on the right y-axis) as a function of anisotropy parameter $\gamma$ in the limit of $\tau \to 0$. A moderate thermalization time ranges from $61$ to $191$ for anisotropy $0 \leq \gamma \leq 1$ when the trace distance $D(\hat{\rho},\hat{\rho}_A)\sim 10^{-2}$. The other parameters are $B_{1} = 1$, $B_{2} = 2$, $T = 1$.}
   \label{fig:Power Vs Anisotropy}
\end{figure}

The isochoric cooling process of the system with a thermal bath is not an instantaneous process, and ideally takes infinite time. To make a power analysis, we assume that the state of the system becomes very close to a thermal state $\rho_A$ at a finite time $t_c$. In order to make an estimate of this closeness we have calculated the trace distance between two states $\hat{\rho}$ and $\hat{\rho}_A $, defined as $
D(\hat{\rho}, \hat{\rho}_A)=\frac{1}{2} \operatorname{Tr}\left|\hat{\rho}-\hat{\rho}_A\right|
$ \cite{camati2019PRA}, where the state $\hat{\rho}$ is obtained by solving by the master equation [\textbf{Eq.~(\ref{master}}], with time-independent dissipation rate coefficients, as the magnetic field is kept constant at $B=B_1$ during this cooling process. Also, as defined in \textbf{Sec.~\ref{cycle}}, $\rho_A$ represents the thermal state at a temperature $T$ (that of the cold bath) when the magnetic field is maintained at $B=B_1$ (also see \textbf{Fig.~(\ref{fig:schematic diagram}}). We estimated $t_c$, as a time when the trace distance $D(\hat{\rho},\hat{\rho_A})$ becomes $ \sim 10^{-2}$. Such a finite-time analysis of the cooling process helps us in defining the power of the engine as $P = W/(2\tau + t_{c})$, where $\tau$ is the duration of each unitary stage, and we have assumed that the measurement is an instantaneous process. The plot of the power as a function of the anisotropy is shown in \textbf{Fig.~\ref{fig:Power Vs Anisotropy}}.

We found that in the limit of $\tau \to 0$, the work does not change substantially with respect to $\gamma$, as compared to the quasistatic limit, which is discussed in the \textbf{Sec.~\ref{analysis}}. However, the thermalization time increases with the increase in anisotropy, leading to a reduction of power. Further, if we consider that both spins interact with the bath, thermalization of the system during the isochoric cooling process can be achieved at a much faster pace, and hence more power would be generated by the engine.

\section{Conclusions}\label{concl}

We have studied the performance of a measurement-based QOE in a two-spin working system coupled by the Heisenberg anisotropic XY interaction. A non-selective quantum measurement is used to fuel the engine. The non-commuting nature of the Hamiltonian at two different times initiates transitions between the instantaneous energy eigenstates at finite time unitary processes. Furthermore, the instantaneous energy eigenstates do not coincide with the measurement basis states which causes some transition between them. The relevant thermodynamic quantities are calculated in terms of these transition probabilities. 
We found that the efficiency oscillates largely at short times when the two-spin system is coupled by an anisotropic interaction, while for isotropic interaction there is no oscillation. This oscillation in efficiency is explained in terms of interference between different transition probabilities at finite times. It is observed that the oscillation in efficiency dies out as the unitary processes extend for a longer time and eventually the efficiency approaches the quasistatic limit. Thus, proper control of the duration of unitary processes during transient times can lead to higher work output and less heat absorption. 
As a result, a finite-time engine can be more efficient than a quasi-static engine. The efficiency further increases with increasing anisotropy while in the quasistatic limit, it is observed that the performance deteriorates with an increase in anisotropy.

Also, we studied the performance of the HE under the condition of always-on coupling to the heat bath. We found that the presence of the bath has a negligible effect on its performance in a very short time limit. However, for a longer duration of the stages AB and CD, its performance degrades. This is primarily due to the dominance of the bath interaction over external control during these stages.

\appendix

\section{Eigenvectors and eigenvalues for $\gamma = 0$ and the corresponding transition probabilities}\label{Eigenstate isotropic}
In the limit of $\gamma \to 0$, the eigenstates and corresponding eigenvectors of the Hamiltonian $\hat{H}(t)$ take the following forms:
\begin{equation}
  \begin{array}{l}
  \ket{\psi_{0}} = \ket{00}, ~~~~~~~~~~~~~~~~~~~~~~E_{0} = -2B\\
  \ket{\psi_{1}} = \frac{1}{\sqrt{2}}(-\ket{10} + \ket{01}), ~~~~~E_{1} = -2J\\
  \ket{\psi_{2}} = \frac{1}{\sqrt{2}}(\ket{10} + \ket{01}), ~~~~~~~E_{2} = 2J\\
  \ket{\psi_{3}} = \ket{11},~~~~~~~~~~~~~~~~~~~~~E_{3} = 2B.
  \end{array}  
\end{equation}

Clearly, the states $|\psi_{0,3}\rangle$ are no longer entangled, though they differ from the Bell states $|\psi_\pm\rangle$.  Also in this limit, it can be shown using the above eigenstates,  that the transition probabilities as mentioned in the \textbf{Sec.~\ref{tran prob}}, are reduced to $ \delta = \lambda = \chi = 1/2$ and $\xi = 0$, where we  used: $\langle\psi_{0}^{(2)}|\hat{U}(\tau)|\psi_{3}^{(1)}\rangle =$ $\langle\psi_{3}^{(2)}|\hat{U}(\tau)| \psi_{0}^{(1)}\rangle =$ $\langle\psi_{0}^{(1)}|\hat{V}(\tau)| \psi_{3}^{(2)}\rangle = $ $\langle\psi_{3}^{(1)}|\hat{V}(\tau)| \psi_{0}^{(2)}\rangle = 0$, and $\langle\psi_{0}^{(2)}|\hat{U}(\tau)| \psi_{0}^{(1)}\rangle = $ $\langle\psi_{3}^{(1)}|\hat{V}(\tau)| \psi_{3}^{(2)}\rangle = 1$. Therefore, for the isotropic case, the efficiency becomes $\eta = 1 - B_1/B_2$, which does not depend on $\tau$.

\section{Derivation of the internal energies: }\label{ApeA}

\subsection{At A - }

The Hamiltonian at point A of the cycle can be expressed as
$$
\hat{H}_{A} = \hat{H}_{1} = \sum_{i=0}^3 E_{i}^{(1)}|\psi_{i}^{(1)}\rangle\langle\psi_{i}^{(1)}|
$$

where $\{|\psi_{i}^{(1)}\rangle\}$ are the eigenstates of the Hamiltonian $\hat{H}_{1}$. As we consider that the system at A is in thermal equilibrium with the heat bath, the thermal density matrix is given by 
\begin{equation}
  \hat{\rho}_{A} =  \frac{e^{-\beta \hat{H}_{1}}}{Z} = \sum_{i = 0}^3 P_{i}|\psi_{i}^{(1)}\rangle\langle\psi_{i}^{(1)}|  
\end{equation}
where ${P_{i} = e^{-\beta E_{i}^{(1)}}/Z}$ is the thermal occupation probability of the $i$th eigenstate and $Z$ is the relevant partition function. So, the average internal energy at point A is given by 
\begin{equation}
    \langle E_{A}\rangle = \Tr(\hat{H}_{1}\hat{\rho}_{A}) =\sum_{i = 0}^3 P_{i}E_{i}^{(1)} = -4 K_{1} \frac{\sinh 2 K_{1} \beta}{Z}-2 J \frac{\sinh 2J \beta}{Z}.
\end{equation}

\subsection{At B - }

The Hamiltonian at the point B of the cycle can be expressed as 
$$
\hat{H}_{B} = \hat{H}_{2} = \sum_{i=0}^3 E_{i}^{(2)}|\psi_{i}^{(2)}\rangle\langle\psi_{i}^{(2)}|,
$$
where $\{|\psi_{i}^{(2)}\rangle\}$ are the eigenstates of the Hamiltonian $\hat{H}_{2}$. The density matrix at the point B after the unitary process AB can be obtained as 
\begin{equation}
\hat{\rho}_{B} = \hat{U}(\tau)\hat{\rho}_{A}\hat{U}^{\dag}(\tau) = \sum_{i = 0}^3P_{i}\hat{U}(\tau)|\psi_{i}^{(1)}\rangle\langle\psi_{i}^{(1)}|\hat{U}^{\dag}(\tau)\;.
\end{equation}

The average internal energy at the point B can be obtained as
\begin{equation}
\begin{aligned}
&\langle E_{B}\rangle = \Tr(\hat{H}_{2}\hat{\rho}_{B})\\
&= P_{0}E_{0}^{(2)}~\langle\psi_{0}^{(2)}| \hat{U}(\tau)|\psi_{0}^{(1)}\rangle\langle\psi_{0}^{(1)}|\hat{U}^\dag (\tau)|\psi_{0}^{(2)}\rangle ~+~ P_{3}E_{0}^{(2)}~\langle\psi_{0}^{(2)}| \hat{U}(\tau)|\psi_{3}^{(1)}\rangle\langle\psi_{3}^{(1)}|\hat{U}^\dag (\tau)|\psi_{0}^{(2)}\rangle\\
&+~P_1E_{1}^{(2)}~+P_2E_{2}^{(2)}~+~P_{0}E_{3}^{(2)}~{}\langle\psi_{3}^{(2)}|\hat{U}(\tau)|\psi_{0}^{(1)}\rangle\langle\psi_{0}^{(1)}|\hat{U}^\dag (\tau)|\psi_{3}^{(2)}\rangle\\
&~+~P_{3}E_{3}^{(2)}~\langle\psi_{3}^{(2)}|\hat{U}(\tau)|\psi_{3}^{(1)}\rangle\langle\psi_{3}^{(1)}|\hat{U}^\dag(\tau)~|\psi_{3}^{(2)}\rangle\\
& = P_{0}E_{0}^{(2)}(1 - \xi)~+~P_{3}E_{0}^{(2)}\xi~+~ P_{1}E_{1}^{(2)}~+~P_{2}E_{2}^{(2)}~+~P_{0}E_{3}^{(2)}\xi~+~P_{3}E_{3}^{(2)}(1-\xi)\\
& = -4 K_{2} (1 - 2\xi) \frac{ \sinh 2 K_{1} \beta}{Z}-4J \frac{\sinh 2J \beta}{Z}\;,
\end{aligned}
\end{equation}
where we have used the microreversibility condition $|\langle\psi_{0}^{(2)}| \hat{U}(\tau)|\psi_{3}^{(1)}\rangle |^{2} = |\langle\psi_{3}^{(2)}| \hat{U}(\tau)|\psi_{0}^{(1)}\rangle |^{2} = \xi$ (proof is given in the \textbf{App.~\ref{micro}}) and $ |\langle\psi_{0}^{(2)}| \hat{U}(\tau)|\psi_{0}^{(1)}\rangle |^{2} = |\langle\psi_{3}^{(2)}| \hat{U}(\tau)|\psi_{3}^{(1)}\rangle |^{2} = 1 - \xi$. 
In unitary stages for a short time interval $\tau$, nonadiabatic transitions occur between energy eigenstates that are coupled \cite{cherubim2022PRE}. In the present case, such transitions will be induced between the levels $|\psi_{0}\rangle$ and $|\psi_{3}\rangle$. So, the terms like $\langle\psi_{0}^{(2)}| \hat{U}(\tau)|\psi_{1}^{(1)}\rangle$, $\langle\psi_{0}^{(2)}| \hat{U}(\tau)|\psi_{2}^{(1)}\rangle$, $\langle\psi_{3}^{(2)}| \hat{U}(\tau)|\psi_{1}^{(1)}\rangle$ etc. become zero.

\subsection{At C - }
The density matrix after the measurement stage can be written as 
\begin{equation}
\begin{aligned}
&\hat{\rho}_{C} = \sum_{\alpha = 1}^4  \hat{M}_{\alpha}\hat{\rho}_{B}\hat{M}_{\alpha} \\
&= P_{0}\delta|\psi_+\rangle\langle\psi_+|~+~ P_{0} (1 - \delta)|\psi_-\rangle\langle\psi_-|~+~P_{1}|\psi_{1}^{(1)}\rangle\langle\psi_{1}^{(1)}| ~+~P_{2}|\psi_{2}^{(1)}\rangle\langle\psi_{2}^{(1)}|  \\
&+P_{3}(1 - \delta)|\psi_+\rangle\langle\psi_+|~+~ P_{3} \delta|\psi_-\rangle\langle\psi_-|,
\end{aligned}
\end{equation}
where $\hat{M}_{\alpha}^{\dag} = \hat{M}_{\alpha}$ and we have used the microreversibility condition $ |\langle\psi_+|\hat{U}(\tau)|\psi_{0}^{(1)}\rangle|^2  = |\langle\psi_-|\hat{U}(\tau)|\psi_{3}^{(1)}\rangle|^2 = \delta$ (proof can be found in the  \textbf{App.~\ref{micro}})) and $|\langle\psi_-|\hat{U}(\tau)|\psi_{0}^{(1)}\rangle|^2  = |\langle\psi_+|\hat{U}(\tau)|\psi_{3}^{(1)}\rangle|^2 = 1 - \delta  $.
The average internal energy at the point C can be obtained as 
\begin{equation}
\begin{aligned}
&\langle E_{C}\rangle = \Tr(\hat{H}_{2}\hat{\rho}_{C}) \\
&= E_{0}^{(2)}P_{0}\delta|\langle\psi_{0}^{(2)}|\psi_+\rangle|^2~+~E_{0}^{(2)}P_{0}(1 - \delta)|\langle\psi_{0}^{(2)}|\psi_-\rangle|^2~+~E_{0}^{(2)}P_{3}(1 - \delta)|\langle\psi_{0}^{(2)}|\psi_+\rangle|^2\\
&+E_{0}^{(2)}P_{3}\delta|\langle\psi_{0}^{(2)}|\psi_-\rangle|^2~+~E_{1}^{(2)}P_{1}~+~E_{2}^{(2)}P_{2}~+~E_{3}^{(2)}P_{0}\delta|\langle\psi_{3}^{(2)}|\psi_+\rangle|^2~\\
&+~E_{3}^{(2)}P_{3}(1 - \delta)|\langle\psi_{3}^{(2)}|\psi_-\rangle|^2+~E_{3}^{(2)}P_{3}(1 - \delta)|\langle\psi_{3}^{(2)}|\psi_+\rangle|^2~+~E_{3}^{(2)}P_{3}\delta|\langle\psi_{3}^{(2)}|\psi_-\rangle|^2\\
&=-4 K_{2} (1 - 2\delta) (1 - 2\chi) \frac{ \sinh 2 K_{1} \beta}{Z}-4J \frac{\sinh 2J \beta}{Z}\;,
\end{aligned}
\end{equation}
where we have used $ |\langle\psi_{0}^{(2)}|\psi_+\rangle|^2 = |\langle\psi_{3}^{(2)}|\psi_-\rangle|^2 = \chi$ and $ |\langle\psi_{0}^{(2)}|\psi_-\rangle|^2 = |\langle\psi_{3}^{(2)}|\psi_+\rangle|^2 = 1 - \chi$, which can be proved using the conservation of probability 
$$
|\langle\psi_0^{(2)} \mid \psi_{-}\rangle|^2+|\langle\psi_3^{(2)} \mid \psi_{-}\rangle|^2=1.
$$

\subsection{At D - }

The density matrix at point D after the unitary process CD is given by
\begin{equation}
\hat{\rho}_{D} = \hat{V}(\tau)\hat{\rho}_{C}\hat{V}^{\dag}(\tau)\;.
\end{equation}

Similarly to points A, B, and C, we can derive the average internal energy at point D which is given by

\begin{equation}
\langle E_{D}\rangle = \Tr(\hat{H}_{1}\hat{\rho}_{D}) =-4 K_{1} (1 - 2\delta) (1 - 2\lambda) \frac{\sinh 2 K_{1} \beta}{Z}-4J \frac{\sinh 2J \beta}{Z}\;,
\end{equation}
where we have used the microreversibility condition $|\langle\psi_{0}^{(1)}| \hat{V}(\tau)|\psi_+\rangle|^2 = |\langle\psi_{3}^{(1)}| \hat{V}(\tau)|\psi_-\rangle|^2 = \lambda$ (proof is given in the \textbf{App.~\ref{micro}}) and $ |\langle\psi_{3}^{(1)}| \hat{V}(\tau)|\psi_+\rangle|^2 = |\langle\psi_{0}^{(1)}| \hat{V}(\tau)|\psi_-\rangle|^2 = 1 - \lambda.$

\section{Proof of the microreversibility conditions:}\label{micro}
We show below in details the proof of the relation $|\langle\psi_{0}^{(2)}| \hat{U}(\tau)|\psi_{3}^{(1)}\rangle |^{2} = |\langle\psi_{3}^{(2)}| \hat{U}(\tau)|\psi_{0}^{(1)}\rangle |^{2}$.

\begin{eqnarray}
&&|\langle\psi_{3}^{(2)}| \hat{U}(\tau)|\psi_{0}^{(1)}\rangle |^{2} = \langle\psi_{3}^{(2)}| \hat{U}(\tau)|\psi_{0}^{(1)}\rangle\langle\psi_{0}^{(1)}|\hat{U}^{\dag}(\tau)|\psi_{3}^{(2)}\rangle\nonumber\\ 
&&=\langle\psi_{3}^{(2)}| \hat{U}(\tau)(\mathbb{I} - |\psi_{1}^{(1)}\rangle\langle\psi_{1}^{(1)}|~-~|\psi_{2}^{(1)}\rangle\langle\psi_{2}^{(1)}|~-~|\psi_{3}^{(1)}\rangle\langle\psi_{3}^{(1)}|) \hat{U}^{\dag}(\tau)|\psi_{3}^2\rangle \;,
\end{eqnarray}
where we have used the completeness relation $\sum_{i=0}^3|\psi_i^{(1)}\rangle \langle \psi_i^{(1)}| = \mathbb{I}$. The above relation can then be rewritten as \begin{eqnarray}
 &&\langle\psi_{3}^{(2)}|\hat{U}(\tau)\hat{U}^{\dag}(\tau)|\psi_{3}^{(2)}\rangle - |\langle\psi_{3}^{(2)}| \hat{U}(\tau)|\psi_{3}^{(1)}\rangle |^{2} = 1 - \left(1 - |\langle\psi_{0}^{(2)}| \hat{U}(\tau)|\psi_{3}^{(1)}\rangle |^{2}\right)\nonumber\\
 &&= |\langle\psi_{0}^{(2)}| \hat{U}(\tau)|\psi_{3}^{(1)}\rangle |^{2}\;.
 \end{eqnarray}
In the last step of the above derivation, we have used the conservation of the probability: $$|\langle\psi_{0}^{(2)}| \hat{U}(\tau)|\psi_{3}^{(1)}\rangle |^{2}~+~ |\langle\psi_{3}^{(2)}| \hat{U}(\tau)|\psi_{3}^{(1)}\rangle |^{2} = 1.$$

Similarly, we can prove the other microreversibility conditions, namely, $ |\langle\psi_+| \hat{U}(\tau)|\psi_{0}^{(1)}\rangle |^{2} = |\langle\psi_-| \hat{U}(\tau)|\psi_{3}^{(1)}\rangle |^{2}$,
and $|\langle\psi_0^{(1)}|\hat{V}(\tau)| \psi_{+}\rangle|^2=|\langle\psi_3^{(1)}|\hat{V}(\tau)| \psi_{-}\rangle|^2$
by using $\sum_{\alpha=1}^4 \hat{M}_\alpha^2 = \mathds{1}$ and the conservation of probability, respectively
$$
\begin{gathered}
|\langle\psi_+| \hat{U}(\tau)|\psi_{3}^{(1)}\rangle |^{2}~+~ |\langle\psi_-| \hat{U}(\tau)|\psi_{3}^{(1)}\rangle |^{2} = 1,\\
|\langle\psi_0^{(1)}|\hat{V}(\tau)| \psi_{-}\rangle|^2~+~|\langle\psi_3^{(1)}|\hat{V}(\tau)| \psi_{-}\rangle|^2=1.
\end{gathered}
$$

\bibliography{ref}

\begin{thebibliography}{70}
\expandafter\ifx\csname natexlab\endcsname\relax\def\natexlab#1{#1}\fi
\expandafter\ifx\csname bibnamefont\endcsname\relax
  \def\bibnamefont#1{#1}\fi
\expandafter\ifx\csname bibfnamefont\endcsname\relax
  \def\bibfnamefont#1{#1}\fi
\expandafter\ifx\csname citenamefont\endcsname\relax
  \def\citenamefont#1{#1}\fi
\expandafter\ifx\csname url\endcsname\relax
  \def\url#1{\texttt{#1}}\fi
\expandafter\ifx\csname urlprefix\endcsname\relax\def\urlprefix{URL }\fi
\providecommand{\bibinfo}[2]{#2}
\providecommand{\eprint}[2][]{\url{#2}}

\bibitem[{\citenamefont{Latune et~al.}(2021)\citenamefont{Latune, Sinayskiy,
  and Petruccione}}]{latune2021EPJST}
\bibinfo{author}{\bibfnamefont{C.~L.} \bibnamefont{Latune}},
  \bibinfo{author}{\bibfnamefont{I.}~\bibnamefont{Sinayskiy}},
  \bibnamefont{and}
  \bibinfo{author}{\bibfnamefont{F.}~\bibnamefont{Petruccione}},
  \bibinfo{journal}{The European Physical Journal Special Topics}
  \textbf{\bibinfo{volume}{230}}, \bibinfo{pages}{841} (\bibinfo{year}{2021}).

\bibitem[{\citenamefont{Mitchison et~al.}(2015)\citenamefont{Mitchison, Woods,
  Prior, and Huber}}]{mitchison2015NJP}
\bibinfo{author}{\bibfnamefont{M.~T.} \bibnamefont{Mitchison}},
  \bibinfo{author}{\bibfnamefont{M.~P.} \bibnamefont{Woods}},
  \bibinfo{author}{\bibfnamefont{J.}~\bibnamefont{Prior}}, \bibnamefont{and}
  \bibinfo{author}{\bibfnamefont{M.}~\bibnamefont{Huber}},
  \bibinfo{journal}{New Journal of Physics} \textbf{\bibinfo{volume}{17}},
  \bibinfo{pages}{115013} (\bibinfo{year}{2015}).

\bibitem[{\citenamefont{Scully et~al.}(2003)\citenamefont{Scully, Zubairy,
  Agarwal, and Walther}}]{scully2003Science}
\bibinfo{author}{\bibfnamefont{M.~O.} \bibnamefont{Scully}},
  \bibinfo{author}{\bibfnamefont{M.~S.} \bibnamefont{Zubairy}},
  \bibinfo{author}{\bibfnamefont{G.~S.} \bibnamefont{Agarwal}},
  \bibnamefont{and} \bibinfo{author}{\bibfnamefont{H.}~\bibnamefont{Walther}},
  \bibinfo{journal}{Science} \textbf{\bibinfo{volume}{299}},
  \bibinfo{pages}{862} (\bibinfo{year}{2003}).

\bibitem[{\citenamefont{Shi et~al.}(2020)\citenamefont{Shi, Shi, Wang, Hu, Liu,
  Yang, and Fan}}]{shi2020JPA}
\bibinfo{author}{\bibfnamefont{Y.-H.} \bibnamefont{Shi}},
  \bibinfo{author}{\bibfnamefont{H.-L.} \bibnamefont{Shi}},
  \bibinfo{author}{\bibfnamefont{X.-H.} \bibnamefont{Wang}},
  \bibinfo{author}{\bibfnamefont{M.-L.} \bibnamefont{Hu}},
  \bibinfo{author}{\bibfnamefont{S.-Y.} \bibnamefont{Liu}},
  \bibinfo{author}{\bibfnamefont{W.-L.} \bibnamefont{Yang}}, \bibnamefont{and}
  \bibinfo{author}{\bibfnamefont{H.}~\bibnamefont{Fan}},
  \bibinfo{journal}{Journal of Physics A: Mathematical and Theoretical}
  \textbf{\bibinfo{volume}{53}}, \bibinfo{pages}{085301}
  (\bibinfo{year}{2020}).

\bibitem[{\citenamefont{Latune et~al.}(2019)\citenamefont{Latune, Sinayskiy,
  and Petruccione}}]{latune2019SciRep}
\bibinfo{author}{\bibfnamefont{C.}~\bibnamefont{Latune}},
  \bibinfo{author}{\bibfnamefont{I.}~\bibnamefont{Sinayskiy}},
  \bibnamefont{and}
  \bibinfo{author}{\bibfnamefont{F.}~\bibnamefont{Petruccione}},
  \bibinfo{journal}{Scientific Reports} \textbf{\bibinfo{volume}{9}},
  \bibinfo{pages}{1} (\bibinfo{year}{2019}).

\bibitem[{\citenamefont{Brandner et~al.}(2015)\citenamefont{Brandner, Bauer,
  Schmid, and Seifert}}]{brandner2015NJP}
\bibinfo{author}{\bibfnamefont{K.}~\bibnamefont{Brandner}},
  \bibinfo{author}{\bibfnamefont{M.}~\bibnamefont{Bauer}},
  \bibinfo{author}{\bibfnamefont{M.~T.} \bibnamefont{Schmid}},
  \bibnamefont{and} \bibinfo{author}{\bibfnamefont{U.}~\bibnamefont{Seifert}},
  \bibinfo{journal}{New Journal of Physics} \textbf{\bibinfo{volume}{17}},
  \bibinfo{pages}{065006} (\bibinfo{year}{2015}).

\bibitem[{\citenamefont{Uzdin et~al.}(2015)\citenamefont{Uzdin, Levy, and
  Kosloff}}]{uzdin2015PRX}
\bibinfo{author}{\bibfnamefont{R.}~\bibnamefont{Uzdin}},
  \bibinfo{author}{\bibfnamefont{A.}~\bibnamefont{Levy}}, \bibnamefont{and}
  \bibinfo{author}{\bibfnamefont{R.}~\bibnamefont{Kosloff}},
  \bibinfo{journal}{Physical Review X} \textbf{\bibinfo{volume}{5}},
  \bibinfo{pages}{031044} (\bibinfo{year}{2015}).

\bibitem[{\citenamefont{Uzdin}(2016)}]{uzdin2016PRAp}
\bibinfo{author}{\bibfnamefont{R.}~\bibnamefont{Uzdin}},
  \bibinfo{journal}{Physical Review Applied} \textbf{\bibinfo{volume}{6}},
  \bibinfo{pages}{024004} (\bibinfo{year}{2016}).

\bibitem[{\citenamefont{Brandner et~al.}(2017)\citenamefont{Brandner, Bauer,
  and Seifert}}]{brandner2017PRL}
\bibinfo{author}{\bibfnamefont{K.}~\bibnamefont{Brandner}},
  \bibinfo{author}{\bibfnamefont{M.}~\bibnamefont{Bauer}}, \bibnamefont{and}
  \bibinfo{author}{\bibfnamefont{U.}~\bibnamefont{Seifert}},
  \bibinfo{journal}{Physical Review Letters} \textbf{\bibinfo{volume}{119}},
  \bibinfo{pages}{170602} (\bibinfo{year}{2017}).

\bibitem[{\citenamefont{Scully et~al.}(2011)\citenamefont{Scully, Chapin,
  Dorfman, Kim, and Svidzinsky}}]{scully2011PNAS}
\bibinfo{author}{\bibfnamefont{M.~O.} \bibnamefont{Scully}},
  \bibinfo{author}{\bibfnamefont{K.~R.} \bibnamefont{Chapin}},
  \bibinfo{author}{\bibfnamefont{K.~E.} \bibnamefont{Dorfman}},
  \bibinfo{author}{\bibfnamefont{M.~B.} \bibnamefont{Kim}}, \bibnamefont{and}
  \bibinfo{author}{\bibfnamefont{A.}~\bibnamefont{Svidzinsky}},
  \bibinfo{journal}{Proceedings of the National Academy of Sciences}
  \textbf{\bibinfo{volume}{108}}, \bibinfo{pages}{15097}
  (\bibinfo{year}{2011}).

\bibitem[{\citenamefont{Rahav et~al.}(2012)\citenamefont{Rahav, Harbola, and
  Mukamel}}]{rahav2012PRA}
\bibinfo{author}{\bibfnamefont{S.}~\bibnamefont{Rahav}},
  \bibinfo{author}{\bibfnamefont{U.}~\bibnamefont{Harbola}}, \bibnamefont{and}
  \bibinfo{author}{\bibfnamefont{S.}~\bibnamefont{Mukamel}},
  \bibinfo{journal}{Physical Review A} \textbf{\bibinfo{volume}{86}},
  \bibinfo{pages}{043843} (\bibinfo{year}{2012}).

\bibitem[{\citenamefont{Brunner et~al.}(2014)\citenamefont{Brunner, Huber,
  Linden, Popescu, Silva, and Skrzypczyk}}]{brunner2014PRE}
\bibinfo{author}{\bibfnamefont{N.}~\bibnamefont{Brunner}},
  \bibinfo{author}{\bibfnamefont{M.}~\bibnamefont{Huber}},
  \bibinfo{author}{\bibfnamefont{N.}~\bibnamefont{Linden}},
  \bibinfo{author}{\bibfnamefont{S.}~\bibnamefont{Popescu}},
  \bibinfo{author}{\bibfnamefont{R.}~\bibnamefont{Silva}}, \bibnamefont{and}
  \bibinfo{author}{\bibfnamefont{P.}~\bibnamefont{Skrzypczyk}},
  \bibinfo{journal}{Physical Review E} \textbf{\bibinfo{volume}{89}},
  \bibinfo{pages}{032115} (\bibinfo{year}{2014}).

\bibitem[{\citenamefont{Altintas et~al.}(2014)\citenamefont{Altintas, Hardal,
  and M\"ustecapl\ifmmode \imath \else \i \fi{}og\ifmmode~\tilde{}\else
  \~{}\fi{}lu}}]{Altintas2014PRE}
\bibinfo{author}{\bibfnamefont{F.}~\bibnamefont{Altintas}},
  \bibinfo{author}{\bibfnamefont{A.~U.~C.} \bibnamefont{Hardal}},
  \bibnamefont{and} \bibinfo{author}{\bibfnamefont{O.~E.}
  \bibnamefont{M\"ustecapl\ifmmode \imath \else \i
  \fi{}og\ifmmode~\tilde{}\else \~{}\fi{}lu}}, \bibinfo{journal}{Physical
  Review E} \textbf{\bibinfo{volume}{90}}, \bibinfo{pages}{032102}
  (\bibinfo{year}{2014}).

\bibitem[{\citenamefont{Barrios et~al.}(2017)\citenamefont{Barrios,
  Albarr{\'a}n-Arriagada, C{\'a}rdenas-L{\'o}pez, Romero, and
  Retamal}}]{barrios2017PRA}
\bibinfo{author}{\bibfnamefont{G.~A.} \bibnamefont{Barrios}},
  \bibinfo{author}{\bibfnamefont{F.}~\bibnamefont{Albarr{\'a}n-Arriagada}},
  \bibinfo{author}{\bibfnamefont{F.}~\bibnamefont{C{\'a}rdenas-L{\'o}pez}},
  \bibinfo{author}{\bibfnamefont{G.}~\bibnamefont{Romero}}, \bibnamefont{and}
  \bibinfo{author}{\bibfnamefont{J.}~\bibnamefont{Retamal}},
  \bibinfo{journal}{Physical Review A} \textbf{\bibinfo{volume}{96}},
  \bibinfo{pages}{052119} (\bibinfo{year}{2017}).

\bibitem[{\citenamefont{Altintas et~al.}(2015)\citenamefont{Altintas, Hardal,
  and M{\"u}stecapl{\i}o{\u{g}}lu}}]{altintas2015PRA}
\bibinfo{author}{\bibfnamefont{F.}~\bibnamefont{Altintas}},
  \bibinfo{author}{\bibfnamefont{A.~{\"U}.} \bibnamefont{Hardal}},
  \bibnamefont{and} \bibinfo{author}{\bibfnamefont{{\"O}.~E.}
  \bibnamefont{M{\"u}stecapl{\i}o{\u{g}}lu}}, \bibinfo{journal}{Physical Review
  A} \textbf{\bibinfo{volume}{91}}, \bibinfo{pages}{023816}
  (\bibinfo{year}{2015}).

\bibitem[{\citenamefont{Hewgill et~al.}(2018)\citenamefont{Hewgill, Ferraro,
  and De~Chiara}}]{hewgill2018PRA}
\bibinfo{author}{\bibfnamefont{A.}~\bibnamefont{Hewgill}},
  \bibinfo{author}{\bibfnamefont{A.}~\bibnamefont{Ferraro}}, \bibnamefont{and}
  \bibinfo{author}{\bibfnamefont{G.}~\bibnamefont{De~Chiara}},
  \bibinfo{journal}{Physical Review A} \textbf{\bibinfo{volume}{98}},
  \bibinfo{pages}{042102} (\bibinfo{year}{2018}).

\bibitem[{\citenamefont{Zhang et~al.}(2007)\citenamefont{Zhang, Liu, Chen, and
  Li}}]{zhang2007PRA}
\bibinfo{author}{\bibfnamefont{T.}~\bibnamefont{Zhang}},
  \bibinfo{author}{\bibfnamefont{W.-T.} \bibnamefont{Liu}},
  \bibinfo{author}{\bibfnamefont{P.-X.} \bibnamefont{Chen}}, \bibnamefont{and}
  \bibinfo{author}{\bibfnamefont{C.-Z.} \bibnamefont{Li}},
  \bibinfo{journal}{Physical Review A} \textbf{\bibinfo{volume}{75}},
  \bibinfo{pages}{062102} (\bibinfo{year}{2007}).

\bibitem[{\citenamefont{Das and Ghosh}(2019)}]{das2019Entropy}
\bibinfo{author}{\bibfnamefont{A.}~\bibnamefont{Das}} \bibnamefont{and}
  \bibinfo{author}{\bibfnamefont{S.}~\bibnamefont{Ghosh}},
  \bibinfo{journal}{Entropy} \textbf{\bibinfo{volume}{21}},
  \bibinfo{pages}{1131} (\bibinfo{year}{2019}).

\bibitem[{\citenamefont{Thomas and Johal}(2011)}]{thomas2011PRE}
\bibinfo{author}{\bibfnamefont{G.}~\bibnamefont{Thomas}} \bibnamefont{and}
  \bibinfo{author}{\bibfnamefont{R.~S.} \bibnamefont{Johal}},
  \bibinfo{journal}{Physical Review E} \textbf{\bibinfo{volume}{83}},
  \bibinfo{pages}{031135} (\bibinfo{year}{2011}).

\bibitem[{\citenamefont{{\c{C}}akmak et~al.}(2016)\citenamefont{{\c{C}}akmak,
  Altintas, and E~M{\"u}stecapl{\i}o{\u{g}}lu}}]{ccakmak2016EPJP}
\bibinfo{author}{\bibfnamefont{S.}~\bibnamefont{{\c{C}}akmak}},
  \bibinfo{author}{\bibfnamefont{F.}~\bibnamefont{Altintas}}, \bibnamefont{and}
  \bibinfo{author}{\bibfnamefont{{\"O}.}~\bibnamefont{E~M{\"u}stecapl{\i}o{\u{g}}lu}},
  \bibinfo{journal}{The European Physical Journal Plus}
  \textbf{\bibinfo{volume}{131}}, \bibinfo{pages}{1} (\bibinfo{year}{2016}).

\bibitem[{\citenamefont{{\c{C}}akmak
  et~al.}(2017{\natexlab{a}})\citenamefont{{\c{C}}akmak, T{\"u}rkpen{\c{c}}e,
  and Altintas}}]{ccakmak2017EPJP}
\bibinfo{author}{\bibfnamefont{S.}~\bibnamefont{{\c{C}}akmak}},
  \bibinfo{author}{\bibfnamefont{D.}~\bibnamefont{T{\"u}rkpen{\c{c}}e}},
  \bibnamefont{and} \bibinfo{author}{\bibfnamefont{F.}~\bibnamefont{Altintas}},
  \bibinfo{journal}{The European Physical Journal Plus}
  \textbf{\bibinfo{volume}{132}}, \bibinfo{pages}{1}
  (\bibinfo{year}{2017}{\natexlab{a}}).

\bibitem[{\citenamefont{Altintas and
  M{\"u}stecapl{\i}o{\u{g}}lu}(2015)}]{altintas2015PRE}
\bibinfo{author}{\bibfnamefont{F.}~\bibnamefont{Altintas}} \bibnamefont{and}
  \bibinfo{author}{\bibfnamefont{{\"O}.~E.}
  \bibnamefont{M{\"u}stecapl{\i}o{\u{g}}lu}}, \bibinfo{journal}{Physical Review
  E} \textbf{\bibinfo{volume}{92}}, \bibinfo{pages}{022142}
  (\bibinfo{year}{2015}).

\bibitem[{\citenamefont{Ivanchenko}(2015)}]{ivanchenko2015PRE}
\bibinfo{author}{\bibfnamefont{E.}~\bibnamefont{Ivanchenko}},
  \bibinfo{journal}{Physical Review E} \textbf{\bibinfo{volume}{92}},
  \bibinfo{pages}{032124} (\bibinfo{year}{2015}).

\bibitem[{\citenamefont{Huang et~al.}(2020{\natexlab{a}})\citenamefont{Huang,
  Yang, Zhang, Zhao, and Wu}}]{huang2020QIP}
\bibinfo{author}{\bibfnamefont{X.-L.} \bibnamefont{Huang}},
  \bibinfo{author}{\bibfnamefont{A.}~\bibnamefont{Yang}},
  \bibinfo{author}{\bibfnamefont{H.}~\bibnamefont{Zhang}},
  \bibinfo{author}{\bibfnamefont{S.}~\bibnamefont{Zhao}}, \bibnamefont{and}
  \bibinfo{author}{\bibfnamefont{S.}~\bibnamefont{Wu}},
  \bibinfo{journal}{Quantum Information Processing}
  \textbf{\bibinfo{volume}{19}}, \bibinfo{pages}{1}
  (\bibinfo{year}{2020}{\natexlab{a}}).

\bibitem[{\citenamefont{Huang et~al.}(2014)\citenamefont{Huang, Liu, Wang, and
  Niu}}]{huang2014EPJP}
\bibinfo{author}{\bibfnamefont{X.}~\bibnamefont{Huang}},
  \bibinfo{author}{\bibfnamefont{Y.}~\bibnamefont{Liu}},
  \bibinfo{author}{\bibfnamefont{Z.}~\bibnamefont{Wang}}, \bibnamefont{and}
  \bibinfo{author}{\bibfnamefont{X.}~\bibnamefont{Niu}}, \bibinfo{journal}{The
  European Physical Journal Plus} \textbf{\bibinfo{volume}{129}},
  \bibinfo{pages}{1} (\bibinfo{year}{2014}).

\bibitem[{\citenamefont{De~Chiara and Antezza}(2020)}]{de2020PRR}
\bibinfo{author}{\bibfnamefont{G.}~\bibnamefont{De~Chiara}} \bibnamefont{and}
  \bibinfo{author}{\bibfnamefont{M.}~\bibnamefont{Antezza}},
  \bibinfo{journal}{Physical Review Research} \textbf{\bibinfo{volume}{2}},
  \bibinfo{pages}{033315} (\bibinfo{year}{2020}).

\bibitem[{\citenamefont{Huang et~al.}(2012)\citenamefont{Huang, Wang, Yi
  et~al.}}]{huang2012PRE}
\bibinfo{author}{\bibfnamefont{X.}~\bibnamefont{Huang}},
  \bibinfo{author}{\bibfnamefont{T.}~\bibnamefont{Wang}},
  \bibinfo{author}{\bibfnamefont{X.}~\bibnamefont{Yi}}, \bibnamefont{et~al.},
  \bibinfo{journal}{Physical Review E} \textbf{\bibinfo{volume}{86}},
  \bibinfo{pages}{051105} (\bibinfo{year}{2012}).

\bibitem[{\citenamefont{Ro{\ss}nagel et~al.}(2014)\citenamefont{Ro{\ss}nagel,
  Abah, Schmidt-Kaler, Singer, and Lutz}}]{rossnagel2014PRL}
\bibinfo{author}{\bibfnamefont{J.}~\bibnamefont{Ro{\ss}nagel}},
  \bibinfo{author}{\bibfnamefont{O.}~\bibnamefont{Abah}},
  \bibinfo{author}{\bibfnamefont{F.}~\bibnamefont{Schmidt-Kaler}},
  \bibinfo{author}{\bibfnamefont{K.}~\bibnamefont{Singer}}, \bibnamefont{and}
  \bibinfo{author}{\bibfnamefont{E.}~\bibnamefont{Lutz}},
  \bibinfo{journal}{Physical Review Letters} \textbf{\bibinfo{volume}{112}},
  \bibinfo{pages}{030602} (\bibinfo{year}{2014}).

\bibitem[{\citenamefont{Alicki and Gelbwaser-Klimovsky}(2015)}]{alicki2015NJP}
\bibinfo{author}{\bibfnamefont{R.}~\bibnamefont{Alicki}} \bibnamefont{and}
  \bibinfo{author}{\bibfnamefont{D.}~\bibnamefont{Gelbwaser-Klimovsky}},
  \bibinfo{journal}{New Journal of Physics} \textbf{\bibinfo{volume}{17}},
  \bibinfo{pages}{115012} (\bibinfo{year}{2015}).

\bibitem[{\citenamefont{de~Assis et~al.}(2019)\citenamefont{de~Assis,
  de~Mendon{\c{c}}a, Villas-Boas, de~Souza, Sarthour, Oliveira, and
  de~Almeida}}]{de2019PRL}
\bibinfo{author}{\bibfnamefont{R.~J.} \bibnamefont{de~Assis}},
  \bibinfo{author}{\bibfnamefont{T.~M.} \bibnamefont{de~Mendon{\c{c}}a}},
  \bibinfo{author}{\bibfnamefont{C.~J.} \bibnamefont{Villas-Boas}},
  \bibinfo{author}{\bibfnamefont{A.~M.} \bibnamefont{de~Souza}},
  \bibinfo{author}{\bibfnamefont{R.~S.} \bibnamefont{Sarthour}},
  \bibinfo{author}{\bibfnamefont{I.~S.} \bibnamefont{Oliveira}},
  \bibnamefont{and} \bibinfo{author}{\bibfnamefont{N.~G.}
  \bibnamefont{de~Almeida}}, \bibinfo{journal}{Physical Review Letters}
  \textbf{\bibinfo{volume}{122}}, \bibinfo{pages}{240602}
  (\bibinfo{year}{2019}).

\bibitem[{\citenamefont{Scully}(2001)}]{scully2001PRL}
\bibinfo{author}{\bibfnamefont{M.~O.} \bibnamefont{Scully}},
  \bibinfo{journal}{Physical Review Letters} \textbf{\bibinfo{volume}{87}},
  \bibinfo{pages}{220601} (\bibinfo{year}{2001}).

\bibitem[{\citenamefont{Klaers et~al.}(2017)\citenamefont{Klaers, Faelt,
  Imamoglu, and Togan}}]{klaers2017PRX}
\bibinfo{author}{\bibfnamefont{J.}~\bibnamefont{Klaers}},
  \bibinfo{author}{\bibfnamefont{S.}~\bibnamefont{Faelt}},
  \bibinfo{author}{\bibfnamefont{A.}~\bibnamefont{Imamoglu}}, \bibnamefont{and}
  \bibinfo{author}{\bibfnamefont{E.}~\bibnamefont{Togan}},
  \bibinfo{journal}{Physical Review X} \textbf{\bibinfo{volume}{7}},
  \bibinfo{pages}{031044} (\bibinfo{year}{2017}).

\bibitem[{\citenamefont{Niedenzu et~al.}(2018)\citenamefont{Niedenzu,
  Mukherjee, Ghosh, Kofman, and Kurizki}}]{niedenzu2018NCom}
\bibinfo{author}{\bibfnamefont{W.}~\bibnamefont{Niedenzu}},
  \bibinfo{author}{\bibfnamefont{V.}~\bibnamefont{Mukherjee}},
  \bibinfo{author}{\bibfnamefont{A.}~\bibnamefont{Ghosh}},
  \bibinfo{author}{\bibfnamefont{A.~G.} \bibnamefont{Kofman}},
  \bibnamefont{and} \bibinfo{author}{\bibfnamefont{G.}~\bibnamefont{Kurizki}},
  \bibinfo{journal}{Nature Communications} \textbf{\bibinfo{volume}{9}},
  \bibinfo{pages}{1} (\bibinfo{year}{2018}).

\bibitem[{\citenamefont{Maruyama et~al.}(2009)\citenamefont{Maruyama, Nori, and
  Vedral}}]{maruyama2009RMP}
\bibinfo{author}{\bibfnamefont{K.}~\bibnamefont{Maruyama}},
  \bibinfo{author}{\bibfnamefont{F.}~\bibnamefont{Nori}}, \bibnamefont{and}
  \bibinfo{author}{\bibfnamefont{V.}~\bibnamefont{Vedral}},
  \bibinfo{journal}{Review of Modern Physics} \textbf{\bibinfo{volume}{81}},
  \bibinfo{pages}{1} (\bibinfo{year}{2009}).

\bibitem[{\citenamefont{Li et~al.}(2012)\citenamefont{Li, Zou, Li, Shao, and
  Wu}}]{liAP2012}
\bibinfo{author}{\bibfnamefont{H.}~\bibnamefont{Li}},
  \bibinfo{author}{\bibfnamefont{J.}~\bibnamefont{Zou}},
  \bibinfo{author}{\bibfnamefont{J.-G.} \bibnamefont{Li}},
  \bibinfo{author}{\bibfnamefont{B.}~\bibnamefont{Shao}}, \bibnamefont{and}
  \bibinfo{author}{\bibfnamefont{L.-A.} \bibnamefont{Wu}},
  \bibinfo{journal}{Annals of Physics} \textbf{\bibinfo{volume}{327}},
  \bibinfo{pages}{2955} (\bibinfo{year}{2012}).

\bibitem[{\citenamefont{Jordan et~al.}(2019)\citenamefont{Jordan, Elouard, and
  Auff{\`e}ves}}]{jordanQSMF2019}
\bibinfo{author}{\bibfnamefont{A.~N.} \bibnamefont{Jordan}},
  \bibinfo{author}{\bibfnamefont{C.}~\bibnamefont{Elouard}}, \bibnamefont{and}
  \bibinfo{author}{\bibfnamefont{A.}~\bibnamefont{Auff{\`e}ves}},
  \bibinfo{journal}{Quantum Studies: Mathematics and Foundations} pp.
  \bibinfo{pages}{1--13} (\bibinfo{year}{2019}).

\bibitem[{\citenamefont{Chand and Biswas}(2017{\natexlab{a}})}]{chand2017EPL}
\bibinfo{author}{\bibfnamefont{S.}~\bibnamefont{Chand}} \bibnamefont{and}
  \bibinfo{author}{\bibfnamefont{A.}~\bibnamefont{Biswas}},
  \bibinfo{journal}{EPL (Europhysics Letters)} \textbf{\bibinfo{volume}{118}},
  \bibinfo{pages}{60003} (\bibinfo{year}{2017}{\natexlab{a}}).

\bibitem[{\citenamefont{Chand and Biswas}(2017{\natexlab{b}})}]{chand2017PRE}
\bibinfo{author}{\bibfnamefont{S.}~\bibnamefont{Chand}} \bibnamefont{and}
  \bibinfo{author}{\bibfnamefont{A.}~\bibnamefont{Biswas}},
  \bibinfo{journal}{Physical Review E} \textbf{\bibinfo{volume}{95}},
  \bibinfo{pages}{032111} (\bibinfo{year}{2017}{\natexlab{b}}).

\bibitem[{\citenamefont{Chand et~al.}(2021)\citenamefont{Chand, Dasgupta, and
  Biswas}}]{chand2021PRE}
\bibinfo{author}{\bibfnamefont{S.}~\bibnamefont{Chand}},
  \bibinfo{author}{\bibfnamefont{S.}~\bibnamefont{Dasgupta}}, \bibnamefont{and}
  \bibinfo{author}{\bibfnamefont{A.}~\bibnamefont{Biswas}},
  \bibinfo{journal}{Physical Review E} \textbf{\bibinfo{volume}{103}},
  \bibinfo{pages}{032144} (\bibinfo{year}{2021}).

\bibitem[{\citenamefont{Yi et~al.}(2017)\citenamefont{Yi, Talkner, and
  Kim}}]{yiPRE2017}
\bibinfo{author}{\bibfnamefont{J.}~\bibnamefont{Yi}},
  \bibinfo{author}{\bibfnamefont{P.}~\bibnamefont{Talkner}}, \bibnamefont{and}
  \bibinfo{author}{\bibfnamefont{Y.~W.} \bibnamefont{Kim}},
  \bibinfo{journal}{Physical Review E} \textbf{\bibinfo{volume}{96}},
  \bibinfo{pages}{022108} (\bibinfo{year}{2017}).

\bibitem[{\citenamefont{Huang et~al.}(2020{\natexlab{b}})\citenamefont{Huang,
  Yang, Zhang, Zhao, and Wu}}]{huangQIP2020}
\bibinfo{author}{\bibfnamefont{X.-L.} \bibnamefont{Huang}},
  \bibinfo{author}{\bibfnamefont{A.}~\bibnamefont{Yang}},
  \bibinfo{author}{\bibfnamefont{H.}~\bibnamefont{Zhang}},
  \bibinfo{author}{\bibfnamefont{S.}~\bibnamefont{Zhao}}, \bibnamefont{and}
  \bibinfo{author}{\bibfnamefont{S.}~\bibnamefont{Wu}},
  \bibinfo{journal}{Quantum Information Processing}
  \textbf{\bibinfo{volume}{19}}, \bibinfo{pages}{1}
  (\bibinfo{year}{2020}{\natexlab{b}}).

\bibitem[{\citenamefont{Denzler and Lutz}(2020)}]{denzler2020PRR}
\bibinfo{author}{\bibfnamefont{T.}~\bibnamefont{Denzler}} \bibnamefont{and}
  \bibinfo{author}{\bibfnamefont{E.}~\bibnamefont{Lutz}},
  \bibinfo{journal}{Physical Review Research} \textbf{\bibinfo{volume}{2}},
  \bibinfo{pages}{032062} (\bibinfo{year}{2020}).

\bibitem[{\citenamefont{Dann et~al.}(2020)\citenamefont{Dann, Kosloff, and
  Salamon}}]{dann2020Entropy}
\bibinfo{author}{\bibfnamefont{R.}~\bibnamefont{Dann}},
  \bibinfo{author}{\bibfnamefont{R.}~\bibnamefont{Kosloff}}, \bibnamefont{and}
  \bibinfo{author}{\bibfnamefont{P.}~\bibnamefont{Salamon}},
  \bibinfo{journal}{Entropy} \textbf{\bibinfo{volume}{22}},
  \bibinfo{pages}{1255} (\bibinfo{year}{2020}).

\bibitem[{\citenamefont{Rezek and Kosloff}(2006)}]{rezek2006NJP}
\bibinfo{author}{\bibfnamefont{Y.}~\bibnamefont{Rezek}} \bibnamefont{and}
  \bibinfo{author}{\bibfnamefont{R.}~\bibnamefont{Kosloff}},
  \bibinfo{journal}{New Journal of Physics} \textbf{\bibinfo{volume}{8}},
  \bibinfo{pages}{83} (\bibinfo{year}{2006}).

\bibitem[{\citenamefont{Lee et~al.}(2020)\citenamefont{Lee, Ha, Park, and
  Jeong}}]{lee2020PRE}
\bibinfo{author}{\bibfnamefont{S.}~\bibnamefont{Lee}},
  \bibinfo{author}{\bibfnamefont{M.}~\bibnamefont{Ha}},
  \bibinfo{author}{\bibfnamefont{J.-M.} \bibnamefont{Park}}, \bibnamefont{and}
  \bibinfo{author}{\bibfnamefont{H.}~\bibnamefont{Jeong}},
  \bibinfo{journal}{Physical Review E} \textbf{\bibinfo{volume}{101}},
  \bibinfo{pages}{022127} (\bibinfo{year}{2020}).

\bibitem[{\citenamefont{Camati et~al.}(2019)\citenamefont{Camati, Santos, and
  Serra}}]{camati2019PRA}
\bibinfo{author}{\bibfnamefont{P.~A.} \bibnamefont{Camati}},
  \bibinfo{author}{\bibfnamefont{J.~F.} \bibnamefont{Santos}},
  \bibnamefont{and} \bibinfo{author}{\bibfnamefont{R.~M.} \bibnamefont{Serra}},
  \bibinfo{journal}{Physical Review A} \textbf{\bibinfo{volume}{99}},
  \bibinfo{pages}{062103} (\bibinfo{year}{2019}).

\bibitem[{\citenamefont{{\c{C}}akmak
  et~al.}(2017{\natexlab{b}})\citenamefont{{\c{C}}akmak, Altintas,
  Gen{\c{c}}ten, and M{\"u}stecapl{\i}o{\u{g}}lu}}]{ccakmak2017EPJD}
\bibinfo{author}{\bibfnamefont{S.}~\bibnamefont{{\c{C}}akmak}},
  \bibinfo{author}{\bibfnamefont{F.}~\bibnamefont{Altintas}},
  \bibinfo{author}{\bibfnamefont{A.}~\bibnamefont{Gen{\c{c}}ten}},
  \bibnamefont{and} \bibinfo{author}{\bibfnamefont{{\"O}.~E.}
  \bibnamefont{M{\"u}stecapl{\i}o{\u{g}}lu}}, \bibinfo{journal}{The European
  Physical Journal D} \textbf{\bibinfo{volume}{71}}, \bibinfo{pages}{1}
  (\bibinfo{year}{2017}{\natexlab{b}}).

\bibitem[{\citenamefont{T{\"u}rkpen{\c{c}}e and
  Altintas}(2019)}]{turkpencce2019QIP}
\bibinfo{author}{\bibfnamefont{D.}~\bibnamefont{T{\"u}rkpen{\c{c}}e}}
  \bibnamefont{and} \bibinfo{author}{\bibfnamefont{F.}~\bibnamefont{Altintas}},
  \bibinfo{journal}{Quantum Information Processing}
  \textbf{\bibinfo{volume}{18}}, \bibinfo{pages}{1} (\bibinfo{year}{2019}).

\bibitem[{\citenamefont{{\c{C}}akmak and
  M{\"u}stecapl{\i}o{\u{g}}lu}(2019)}]{ccakmak2019PRE}
\bibinfo{author}{\bibfnamefont{B.}~\bibnamefont{{\c{C}}akmak}}
  \bibnamefont{and} \bibinfo{author}{\bibfnamefont{{\"O}.~E.}
  \bibnamefont{M{\"u}stecapl{\i}o{\u{g}}lu}}, \bibinfo{journal}{Physical Review
  E} \textbf{\bibinfo{volume}{99}}, \bibinfo{pages}{032108}
  (\bibinfo{year}{2019}).

\bibitem[{\citenamefont{Plastina et~al.}(2014)\citenamefont{Plastina, Alecce,
  Apollaro, Falcone, Francica, Galve, Gullo, and Zambrini}}]{plastina2014PRL}
\bibinfo{author}{\bibfnamefont{F.}~\bibnamefont{Plastina}},
  \bibinfo{author}{\bibfnamefont{A.}~\bibnamefont{Alecce}},
  \bibinfo{author}{\bibfnamefont{T.~J.} \bibnamefont{Apollaro}},
  \bibinfo{author}{\bibfnamefont{G.}~\bibnamefont{Falcone}},
  \bibinfo{author}{\bibfnamefont{G.}~\bibnamefont{Francica}},
  \bibinfo{author}{\bibfnamefont{F.}~\bibnamefont{Galve}},
  \bibinfo{author}{\bibfnamefont{N.~L.} \bibnamefont{Gullo}}, \bibnamefont{and}
  \bibinfo{author}{\bibfnamefont{R.}~\bibnamefont{Zambrini}},
  \bibinfo{journal}{Physical Review Letters} \textbf{\bibinfo{volume}{113}},
  \bibinfo{pages}{260601} (\bibinfo{year}{2014}).

\bibitem[{\citenamefont{Rezek}(2010{\natexlab{a}})}]{rezek2010Ent}
\bibinfo{author}{\bibfnamefont{Y.}~\bibnamefont{Rezek}},
  \bibinfo{journal}{Entropy} \textbf{\bibinfo{volume}{12}},
  \bibinfo{pages}{1885} (\bibinfo{year}{2010}{\natexlab{a}}).

\bibitem[{\citenamefont{Dodonov et~al.}(2018)\citenamefont{Dodonov, Valente,
  and Werlang}}]{dodonov2018JPA}
\bibinfo{author}{\bibfnamefont{A.}~\bibnamefont{Dodonov}},
  \bibinfo{author}{\bibfnamefont{D.}~\bibnamefont{Valente}}, \bibnamefont{and}
  \bibinfo{author}{\bibfnamefont{T.}~\bibnamefont{Werlang}},
  \bibinfo{journal}{Journal of Physics A: Mathematical and Theoretical}
  \textbf{\bibinfo{volume}{51}}, \bibinfo{pages}{365302}
  (\bibinfo{year}{2018}).

\bibitem[{\citenamefont{Abiuso and Giovannetti}(2019)}]{abiuso2019PRA}
\bibinfo{author}{\bibfnamefont{P.}~\bibnamefont{Abiuso}} \bibnamefont{and}
  \bibinfo{author}{\bibfnamefont{V.}~\bibnamefont{Giovannetti}},
  \bibinfo{journal}{Physical Review A} \textbf{\bibinfo{volume}{99}},
  \bibinfo{pages}{052106} (\bibinfo{year}{2019}).

\bibitem[{\citenamefont{Lin et~al.}(2021)\citenamefont{Lin, Su, Chen, Chen, and
  Santos}}]{lin2021PRA}
\bibinfo{author}{\bibfnamefont{Z.}~\bibnamefont{Lin}},
  \bibinfo{author}{\bibfnamefont{S.}~\bibnamefont{Su}},
  \bibinfo{author}{\bibfnamefont{J.}~\bibnamefont{Chen}},
  \bibinfo{author}{\bibfnamefont{J.}~\bibnamefont{Chen}}, \bibnamefont{and}
  \bibinfo{author}{\bibfnamefont{J.~F.} \bibnamefont{Santos}},
  \bibinfo{journal}{Physical Review A} \textbf{\bibinfo{volume}{104}},
  \bibinfo{pages}{062210} (\bibinfo{year}{2021}).

\bibitem[{\citenamefont{Cherubim et~al.}(2022)\citenamefont{Cherubim,
  de~Oliveira, and Jonathan}}]{cherubim2022PRE}
\bibinfo{author}{\bibfnamefont{C.}~\bibnamefont{Cherubim}},
  \bibinfo{author}{\bibfnamefont{T.~R.} \bibnamefont{de~Oliveira}},
  \bibnamefont{and} \bibinfo{author}{\bibfnamefont{D.}~\bibnamefont{Jonathan}},
  \bibinfo{journal}{Physical Review E} \textbf{\bibinfo{volume}{105}},
  \bibinfo{pages}{044120} (\bibinfo{year}{2022}).

\bibitem[{\citenamefont{Suzuki et~al.}(2012)\citenamefont{Suzuki, Inoue, and
  Chakrabarti}}]{suzuki2012book}
\bibinfo{author}{\bibfnamefont{S.}~\bibnamefont{Suzuki}},
  \bibinfo{author}{\bibfnamefont{J.-i.} \bibnamefont{Inoue}}, \bibnamefont{and}
  \bibinfo{author}{\bibfnamefont{B.~K.} \bibnamefont{Chakrabarti}},
  \emph{\bibinfo{title}{Quantum Ising phases and transitions in transverse
  Ising models}}, vol. \bibinfo{volume}{862} (\bibinfo{publisher}{Springer},
  \bibinfo{year}{2012}).

\bibitem[{\citenamefont{Kamta and Starace}(2002)}]{kamta2002PRL}
\bibinfo{author}{\bibfnamefont{G.~L.} \bibnamefont{Kamta}} \bibnamefont{and}
  \bibinfo{author}{\bibfnamefont{A.~F.} \bibnamefont{Starace}},
  \bibinfo{journal}{Physical Review Letters} \textbf{\bibinfo{volume}{88}},
  \bibinfo{pages}{107901} (\bibinfo{year}{2002}).

\bibitem[{\citenamefont{Yeo et~al.}(2005)\citenamefont{Yeo, Liu, Lu, and
  Yang}}]{yeo2005JPA}
\bibinfo{author}{\bibfnamefont{Y.}~\bibnamefont{Yeo}},
  \bibinfo{author}{\bibfnamefont{T.}~\bibnamefont{Liu}},
  \bibinfo{author}{\bibfnamefont{Y.-E.} \bibnamefont{Lu}}, \bibnamefont{and}
  \bibinfo{author}{\bibfnamefont{Q.-Z.} \bibnamefont{Yang}},
  \bibinfo{journal}{Journal of Physics A: Mathematical and General}
  \textbf{\bibinfo{volume}{38}}, \bibinfo{pages}{3235} (\bibinfo{year}{2005}).

\bibitem[{\citenamefont{Rezek}(2010{\natexlab{b}})}]{rezek2010Entropy}
\bibinfo{author}{\bibfnamefont{Y.}~\bibnamefont{Rezek}},
  \bibinfo{journal}{Entropy} \textbf{\bibinfo{volume}{12}},
  \bibinfo{pages}{1885} (\bibinfo{year}{2010}{\natexlab{b}}).

\bibitem[{\citenamefont{Johansson et~al.}(2012)\citenamefont{Johansson, Nation,
  and Nori}}]{johansson2012CPC}
\bibinfo{author}{\bibfnamefont{J.~R.} \bibnamefont{Johansson}},
  \bibinfo{author}{\bibfnamefont{P.~D.} \bibnamefont{Nation}},
  \bibnamefont{and} \bibinfo{author}{\bibfnamefont{F.}~\bibnamefont{Nori}},
  \bibinfo{journal}{Computer Physics Communications}
  \textbf{\bibinfo{volume}{183}}, \bibinfo{pages}{1760} (\bibinfo{year}{2012}).

\bibitem[{\citenamefont{Jiao et~al.}(2021)\citenamefont{Jiao, Zhu, He, Ma, and
  Wang}}]{jiao2021PRE}
\bibinfo{author}{\bibfnamefont{G.}~\bibnamefont{Jiao}},
  \bibinfo{author}{\bibfnamefont{S.}~\bibnamefont{Zhu}},
  \bibinfo{author}{\bibfnamefont{J.}~\bibnamefont{He}},
  \bibinfo{author}{\bibfnamefont{Y.}~\bibnamefont{Ma}}, \bibnamefont{and}
  \bibinfo{author}{\bibfnamefont{J.}~\bibnamefont{Wang}},
  \bibinfo{journal}{Physical Review E} \textbf{\bibinfo{volume}{103}},
  \bibinfo{pages}{032130} (\bibinfo{year}{2021}).

\bibitem[{\citenamefont{Piccione et~al.}(2021)\citenamefont{Piccione,
  De~Chiara, and Bellomo}}]{piccione2021PRA}
\bibinfo{author}{\bibfnamefont{N.}~\bibnamefont{Piccione}},
  \bibinfo{author}{\bibfnamefont{G.}~\bibnamefont{De~Chiara}},
  \bibnamefont{and} \bibinfo{author}{\bibfnamefont{B.}~\bibnamefont{Bellomo}},
  \bibinfo{journal}{Physical Review A} \textbf{\bibinfo{volume}{103}},
  \bibinfo{pages}{032211} (\bibinfo{year}{2021}).

\bibitem[{\citenamefont{Kurizki and Kofman}(2022)}]{kurizki2022book}
\bibinfo{author}{\bibfnamefont{G.}~\bibnamefont{Kurizki}} \bibnamefont{and}
  \bibinfo{author}{\bibfnamefont{A.~G.} \bibnamefont{Kofman}},
  \emph{\bibinfo{title}{Thermodynamics and Control of Open Quantum Systems}}
  (\bibinfo{publisher}{Cambridge University Press}, \bibinfo{year}{2022}).

\bibitem[{\citenamefont{Bhattacharjee and Dutta}(2021)}]{bhattacharjee2021EPJB}
\bibinfo{author}{\bibfnamefont{S.}~\bibnamefont{Bhattacharjee}}
  \bibnamefont{and} \bibinfo{author}{\bibfnamefont{A.}~\bibnamefont{Dutta}},
  \bibinfo{journal}{The European Physical Journal B}
  \textbf{\bibinfo{volume}{94}}, \bibinfo{pages}{1} (\bibinfo{year}{2021}).

\bibitem[{\citenamefont{Dann et~al.}(2018)\citenamefont{Dann, Levy, and
  Kosloff}}]{dann2018PRA}
\bibinfo{author}{\bibfnamefont{R.}~\bibnamefont{Dann}},
  \bibinfo{author}{\bibfnamefont{A.}~\bibnamefont{Levy}}, \bibnamefont{and}
  \bibinfo{author}{\bibfnamefont{R.}~\bibnamefont{Kosloff}},
  \bibinfo{journal}{Physical Review A} \textbf{\bibinfo{volume}{98}},
  \bibinfo{pages}{052129} (\bibinfo{year}{2018}).

\bibitem[{\citenamefont{Leggett et~al.}(1987)\citenamefont{Leggett,
  Chakravarty, Dorsey, Fisher, Garg, and Zwerger}}]{leggett1987RMP}
\bibinfo{author}{\bibfnamefont{A.~J.} \bibnamefont{Leggett}},
  \bibinfo{author}{\bibfnamefont{S.}~\bibnamefont{Chakravarty}},
  \bibinfo{author}{\bibfnamefont{A.~T.} \bibnamefont{Dorsey}},
  \bibinfo{author}{\bibfnamefont{M.~P.} \bibnamefont{Fisher}},
  \bibinfo{author}{\bibfnamefont{A.}~\bibnamefont{Garg}}, \bibnamefont{and}
  \bibinfo{author}{\bibfnamefont{W.}~\bibnamefont{Zwerger}},
  \bibinfo{journal}{Reviews of Modern Physics} \textbf{\bibinfo{volume}{59}},
  \bibinfo{pages}{1} (\bibinfo{year}{1987}).

\bibitem[{\citenamefont{Albash et~al.}(2012)\citenamefont{Albash, Boixo, Lidar,
  and Zanardi}}]{albash2012NJP}
\bibinfo{author}{\bibfnamefont{T.}~\bibnamefont{Albash}},
  \bibinfo{author}{\bibfnamefont{S.}~\bibnamefont{Boixo}},
  \bibinfo{author}{\bibfnamefont{D.~A.} \bibnamefont{Lidar}}, \bibnamefont{and}
  \bibinfo{author}{\bibfnamefont{P.}~\bibnamefont{Zanardi}},
  \bibinfo{journal}{New Journal of Physics} \textbf{\bibinfo{volume}{14}},
  \bibinfo{pages}{123016} (\bibinfo{year}{2012}).

\bibitem[{\citenamefont{Breuer et~al.}(2002)\citenamefont{Breuer, Petruccione
  et~al.}}]{breuer2002book}
\bibinfo{author}{\bibfnamefont{H.-P.} \bibnamefont{Breuer}},
  \bibinfo{author}{\bibfnamefont{F.}~\bibnamefont{Petruccione}},
  \bibnamefont{et~al.}, \emph{\bibinfo{title}{The theory of open quantum
  systems}} (\bibinfo{publisher}{Oxford University Press on Demand},
  \bibinfo{year}{2002}).

\bibitem[{\citenamefont{Hu et~al.}(2018)\citenamefont{Hu, Man, and
  Xia}}]{hu2018QIP}
\bibinfo{author}{\bibfnamefont{L.-Z.} \bibnamefont{Hu}},
  \bibinfo{author}{\bibfnamefont{Z.-X.} \bibnamefont{Man}}, \bibnamefont{and}
  \bibinfo{author}{\bibfnamefont{Y.-J.} \bibnamefont{Xia}},
  \bibinfo{journal}{Quantum Information Processing}
  \textbf{\bibinfo{volume}{17}}, \bibinfo{pages}{1} (\bibinfo{year}{2018}).

\bibitem[{\citenamefont{Liao et~al.}(2011)\citenamefont{Liao, Huang, Kuang
  et~al.}}]{liao2011PRA}
\bibinfo{author}{\bibfnamefont{J.-Q.} \bibnamefont{Liao}},
  \bibinfo{author}{\bibfnamefont{J.-F.} \bibnamefont{Huang}},
  \bibinfo{author}{\bibfnamefont{L.-M.} \bibnamefont{Kuang}},
  \bibnamefont{et~al.}, \bibinfo{journal}{Physical Review A}
  \textbf{\bibinfo{volume}{83}}, \bibinfo{pages}{052110}
  (\bibinfo{year}{2011}).

\end{thebibliography}

\end{document}